\documentclass[aps,prl,reprint,showpacs,showkeys,groupedaddress,nofootinbib,superscriptaddress]{revtex4-1}
\usepackage{hyperref}
\usepackage[T1]{fontenc}
\usepackage[ansinew]{inputenc}
\usepackage{amsmath}
\usepackage{setspace}
\usepackage{psfrag}
\usepackage{mathtools}
\usepackage{amssymb,lineno,amsfonts}
\usepackage{verbatim}
\usepackage{color}
\usepackage{textcomp}
\usepackage[toc,page]{appendix}
\usepackage{graphicx}
\usepackage[caption=false]{subfig}
\usepackage{float}
\usepackage{dblfloatfix}
\usepackage{fixltx2e}
\usepackage{bm}
\usepackage{dcolumn}
\usepackage{gensymb}
\usepackage{ifpdf}
\usepackage{bbm}
\usepackage{mathrsfs}
\usepackage{upgreek}
\usepackage{epstopdf}
\usepackage{esvect}
\usepackage[usenames,dvipsnames]{xcolor}
\definecolor{med-blue}{RGB}{25,25,112}
\hypersetup{colorlinks, linkcolor={blue},citecolor={blue}, urlcolor={blue}}
\usepackage[english]{babel}
\usepackage[autostyle, english = american]{csquotes}

\usepackage{amsfonts}
\usepackage{amssymb}
\begin{document}

\title{Spin domains in ground state of a trapped spin-1 condensate: A general study under Thomas-Fermi approximation}

\author{Projjwal K. Kanjilal}
  \email{projjwal.kanjilal@students.iiserpune.ac.in}
 \affiliation{Indian Institute of Science Education and Research,Pune}
\author{A. Bhattacharyay}%
 \email{a.bhattacharyay@iiserpune.ac.in}
\affiliation{Indian Institute of Science Education and Research,Pune}

\date{\today}

\begin{abstract}
Investigation of ground state structures and phase separation under confinement is of great interest in spinor Bose Einstein Condensates (BEC). In this paper we show that, in general, within the Thomas-Fermi (T-F) approximation, the phase separation scenario of stationary states can be obtained including all the mixed states on an equal footing for a spin-1 condensate for any confinement. Exact analytical expressions of energy density, being independent of local mass density for all allowed states enables this general analysis under T-F approximation. We study here in details a particular case of spherically symmetric harmonic confinement as an example and show a wide range of potential phase separation scenario for anti-ferromagnetic and ferromagnetic interactions.    

\end{abstract}

\maketitle


\section{1.Introduction}\label{section}

 Phase separation of multicomponent Bose-Einstein condensate (BEC) under trapping, as opposed to the phase separation which does not require external fields, was theoretically investigated by Timmermans \cite{PhysRevLett.81.5718} who named it "potential separation". Spin domain formation in an optically trapped sodium spinor condensate has been reported by Stenger et. al., \cite{stenger98} followed by a detailed theoretical justification by Isoshima et. al., \cite{PhysRevA.60.4857}.  A number of theoretical investigations have followed since then to understand the spin domain formation of trapped spin-1 condensate in many different ways \cite{mats10gr,PhysRevA.78.023632,PhysRevA.80.023602,PhysRevA.85.023601}, even at zero magnetic field \cite{PhysRevA.92.023616,jimenez2018spontaneous}. T.-L Ho and V.B Shenoy gave a detailed picture of binary condensates, for which phase separation arises due to the interplay between intra- and inter-species interaction \cite{PhysRevLett.77.3276}. This led to a lot of scientific interest to explore many possible scenarios of domain formation for binary condensates \cite{sabbatini2011phase,vidanovic2013spin,gautam2011phase,PhysRevA.94.013602,doi:10.1063/1.3243875,PhysRevA.82.033609,doi:10.1142/S0217984917502153,PhysRevA.85.043602,PhysRevA.84.013619,PhysRevA.96.043603}. In recent years a lot of thorough scientific investigation provided a detailed picture of instability induced phase separation in a spin orbit coupled condensate \cite{PhysRevLett.105.160403,gautam2014phase,li2018phase,2399-6528-2-2-025008}.
\par
To find out the spin domain formation in the ground state of a spinor BEC, Thomas-Fermi (T-F) approximation is extensively used \cite{PhysRevA.60.4857,PhysRevA.85.043602,PhysRevLett.77.3276,PhysRevLett.80.2027,PhysRevLett.81.742} where the spatial derivatives of order parameter are neglected. This is a  reasonable first step to understand phase separation under entrapment when the trap size is bigger than the healing length \cite{mats10gr}. This procedure provides a wider picture of all possibilities out of which some scenarios might not be present due to instabilities arising from various conditions. However, irrespective of the presence of these instabilities of the stationary states, as a first step, getting a complete picture of coexisting stationary phases in the ground state is desirable. In this paper we follow the T-F approximation to exhaustively investigate the possible phase separations of stationary states under confinement. We show here that, actually, the T-F approximation allows for an exact expression of the energy density of all the possible stationary phases in terms of confining potential and the parameters of the system. This allows for a direct comparison of energy densities of all possible phases on an equal footing at a constant chemical potential to determine which phase is locally of the lowest energy. This becomes possible under the T-F approximation because the energy density can be written as a function of the local total density of the system irrespective of particular phases present. That is, this procedure works equally for all mixed phases of the system. To our knowledge such a comprehensive analysis of the phase separation scenarios under T-F approximation is not in existence yet, however, various specific cases have been discussed under the same approximation. We do the present analysis under the minimal essential constraint of constant chemical potential, however, this exhaustive template would prove useful in understanding phase separation under confinement in a unified way with the possibility of incorporating other constraints.
\par 
On the basis of exact calculations we show here that T-F approximation produces some interesting results. In the presence of anti-ferromagnetic interactions the potential phase separation does not only involve anti-ferromagnetic phases, but also indicate domain formation involving ferromagnetic stationary phases. Three-phase domain formation is only observed when the interactions are anti-ferromagnetic. When the spin-spin interaction is ferromagnetic, under T-F approximation, there actually appears no domain formation involving ferromagnetic phase over the very wide moderately large parameter space that we have explored under isotropic harmonic confinement.  Rather the anti-ferromagnetic and polar phases dominate along with the phase-matched and anti-phase-matched (1,1,1) phases. However, at Zeeman coupling more than $\pm150 Hz$, ferromagnetic phase starts dominating. The (1,1,1) phase indicates presence of all the spin components and would be seen to dominate quite a lot of the domain formation scenarios along with the mixed phases (0,1,1) and (1,1,0). In this paper we show a lot of domain formation possibilities involving these mixed phases following the same basic method of analysis which are not that much reported in the existing literature.
\par
Our present analysis is quite general in terms of consideration of the trapping potential $U(\vec{r})$. We show the domain structures here for a special case of the $U(\vec{r})$, an isotropic harmonic confinement $U(r)=\frac{1}{2}\omega r^2$. However, the same analysis can be used to any potential and can be extended to 2-dimensional or 1-dimensional confinements. The chemical potentials of the basic Zeeman components (1,0,0), (0,1,0) and (0,0,1) are constrained to remain constant for the chemical stability of the co-existing domains and the mixed states. This is a minimal condition, that has to be strictly adhered to in the analysis of phase co-existence. For anti-ferromagnetic and ferromagnetic cases we fix parameters corresponding to $^{23}Na$ and $^{87}Rb$ respectively \cite{PhysRevLett.80.2027,PhysRevLett.87.010404}. 
\par
The plan of the paper is as follows. We begin with the description of the standard mean field analysis using Gross-Pitaevskii equation for a spin-1 BEC and reproduce the phase diagrams of the unconfined case following standard literature. Then we first show the phase separations in the confined case where the spin-spin interaction is negligible and compare results with the unconfined case. A detailed description of phase separation for the anti-ferromagnetic and the ferromagnetic cases follow in the next subsections. We then present a discussion where a comparison of our results under harmonic confinement is compared with the existing ones.

\section{2.Mean field Dynamics of the condensate}\label{mf}
The dynamics of spin-1 condensate under mean field approximation is given by Gross-Pitaevskii (GP) equation \cite{KAWAGUCHI2012253,PhysRevLett.81.742},
\begin{equation}\label{eq:rev1}
i \hbar\dfrac{\partial \psi_m}{\partial t}= \left(\mathcal{H}-pm + qm^2\right)\psi_m
+c_1 \sum_{m'=-1}^1 \vec{F}.\vec{f}_{mm'} \psi_{m'},
\end{equation} 
where the $\mathcal{H}$ corresponds to the symmetric part of the Hamiltonian, $\mathcal{H}=-\dfrac{\hbar^2 \nabla^2}{2M} + U(\vec{r})+c_0 n$ and the suffix $m$ and $m'$ run from $-1$ to $1$ in integer steps. The spin matrices are given by,
$$
  f_x=\frac{1}{\sqrt{2}}
  \begin{bmatrix}
    0 & 1 & 0 \\
    1 & 0 & 1 \\
    0 & 1 & 0
  \end{bmatrix},
  f_y=\frac{i}{\sqrt{2}}
  \begin{bmatrix}
    0 & -1 & 0 \\
    1 & 0 & -1 \\
    0 & 1 & 0
  \end{bmatrix},
    f_z=
  \begin{bmatrix}
    1 & 0 & 0 \\
    0 & 0 & 0 \\
    0 & 0 & -1  
  \end{bmatrix}.
$$\\
\ \ $\psi_m$ is the order parameter corresponding to the $m^{th}$ spin component and $|\psi_m|^2=n_{m}$, gives the density of corresponding spin component. The total density $n=n_1+n_0+n_{-1}$ is the constraint existing everywhere in all that follows. $U(\vec{r})$ is in general, a three dimensional trapping potential and $M$ is the mass of a boson. In the present paper, as a particular case, we will consider 3-dimensional structures of co-existing phases in a condensate trapped by a 3-dimensional harmonic potential, however, our analysis is general. The procedure adopted here can be reduced to 2-dimensional and 1-dimensional confined condensates quite easily by integrating out the extra coordinate(s). If one does that, then coupling constants for the effective 2 or 1-dimensional condensate will get modified with the introduction of the confining length scales \cite{PhysRevA.92.023616}. The parameter $p$ sets the strength of the Zeeman term where, $p=-g\mu_BB$. Here $g$ is the Lande hyperfine $g$-factor, $\mu_B$ is the Bohr magnetron and the magnetic field is applied along the $z$ axis (say) to lift the degeneracy of the spin states. The parameter $q$ is the strength of the quadratic Zeeman term where, $q=(g\mu_BB)^2/\Delta E_{hf}$ with $\Delta E_{hf}$ being the hyperfine splitting. In the above equation, $\vec{F}$ is local spin density vector defined as, \\
\begin{equation}\label{eq:rev2}
F_l(\vec{r}) = \sum_{m,m^\prime=-1}^1{\psi_m^*(\vec{r})(f_l)_{mm^\prime}\psi_{m^\prime}(\vec{r})},
\end{equation}
where $l = x,y,z$. It can be understood that the coefficient of the linear Zeeman term $p$ can include the additive Lagrange multiplier arising from the conservation of magnetization which might be there due to the presence of a magnetic field and total spin orientation conserving scattering. 
The constants $c_1=\dfrac{4\pi \hbar^2}{M} \dfrac{(a_2-a_0)}{3},\quad
c_0= \dfrac{4\pi \hbar^2}{M} \dfrac{(2a_2+a_0)}{3}$ where $a_0$ and  $a_2$ are the s wave scattering lengths for hyperfine spin channels 0 and 2 respectively. Typical values of these scattering lengths in atomic units for $^{23}Na$ $a_2=52.98\pm0.40$ $a.u$, $a_0=47.36\pm0.80$ $a.u$ and for $^{87}Rb$ are $a_2=100.40\pm0.10$ $a.u$, $a_0=101.8\pm0.20$ $a.u$.\cite{KAWAGUCHI2012253}. In what follows, these typical values will be used for ferro- and anti-ferromagnetic cases of analysis.\\

More explicitly, the components of the spin density vectors are,
\begin{equation}\label{eq:rev3}
F_x=\dfrac{1}{\sqrt{2}}\left[\psi^*_{-1}\psi_0+\psi^*_{1}\psi_0+\psi^*_{0}(\psi_{-1}+\psi_1)\right]
\end{equation}

\begin{equation}\label{eq:rev4}
F_y=\dfrac{i}{\sqrt{2}}\left[\psi^*_{-1}\psi_0-\psi^*_{1}\psi_0+\psi^*_{0}(\psi_1-\psi_{-1})\right]
\end{equation}

\begin{equation}\label{eq:rev5}
F_z=\psi^*_{1}\psi_1-\psi^*_{-1}\psi_{-1}
\end{equation}

The mean-field energy of this system can always be written as,
 \begin{equation}\label{eq:rev6}
E[\psi]= \left\langle \hat{H}\right\rangle_0=\int d\vec{r}e(\vec{r}),
\end{equation}
where the local energy density $e(\vec{r})$ is the central quantity which will determine the phase diagrams for a confined system. Explicit expression of the local energy density would read as,
\begin{equation}\label{eq:rev7}
\begin{split}
e(\vec{r})=\sum^1_{m=-1} \psi_m^* &\left[ -\dfrac{\hbar^2 \nabla^2}{2M} + U_{trap}(\vec{r}) -pm + qm^2  \right]\psi_m\\
 &\qquad\qquad\qquad+ \dfrac{c_0}{2} n^2+\dfrac{c_1}{2} |\vec{F}|^2.
\end{split}
\end{equation} 
\par
A Detailed phase diagram of the free system i.e. $U(\vec{r})=0$ is well described in the review \cite{KAWAGUCHI2012253} where going by the ansatz,
\begin{equation}\label{eq:rev8}
\psi_m(\vec{r},t)=\sqrt{n}\zeta_m e^{-i\mu t/\hbar},
\end{equation}
setting $\zeta_0$ real and $Im(\zeta_{+1})=Im(\zeta_{-1})$ by fixing of the overall phase, the following phase diagrams were arrived in ref~\cite{KAWAGUCHI2012253}. In the diagrams shown below, we have used a nomenclature to mark different phases using binary notation of 0 and 1 respectively, meaning zero and non-zero population for particular spin projections. We will be sticking to this notation in what follows because the local density constraint being imposed everywhere, explicit mention of the densities will not be required. However, our following analysis will clearly show that, under T-F approximation, local density of each and every spin state can be found out for their stationary configurations.
\par

These phase diagrams capture five distinct phases separated by boundaries which are function of the parameters $p$, $q$, $c_1$ and the density $n$ of the condensate in presence of a magnetic field. These mean field phase diagrams have been immensely useful in understanding many experimental results \cite{stenger98}. These diagrams, practically at the zero temperature of the condensate indicate a set of (quantum) phase transition boundaries as a function of density. For a trapped BEC, the constant density condition underlying the analysis of a free condensate is no longer valid. Phase separation, therefore, can arise in a trapped spinor condensate which we are going to look at systematically in the following to capture a complete and coherent mean-field description.
\begin{widetext}
	\begin{figure}[h!]
		\subfloat[$c_1>0$\label{subfig-1:hom1}]{%
			\includegraphics[width=0.33\textwidth]{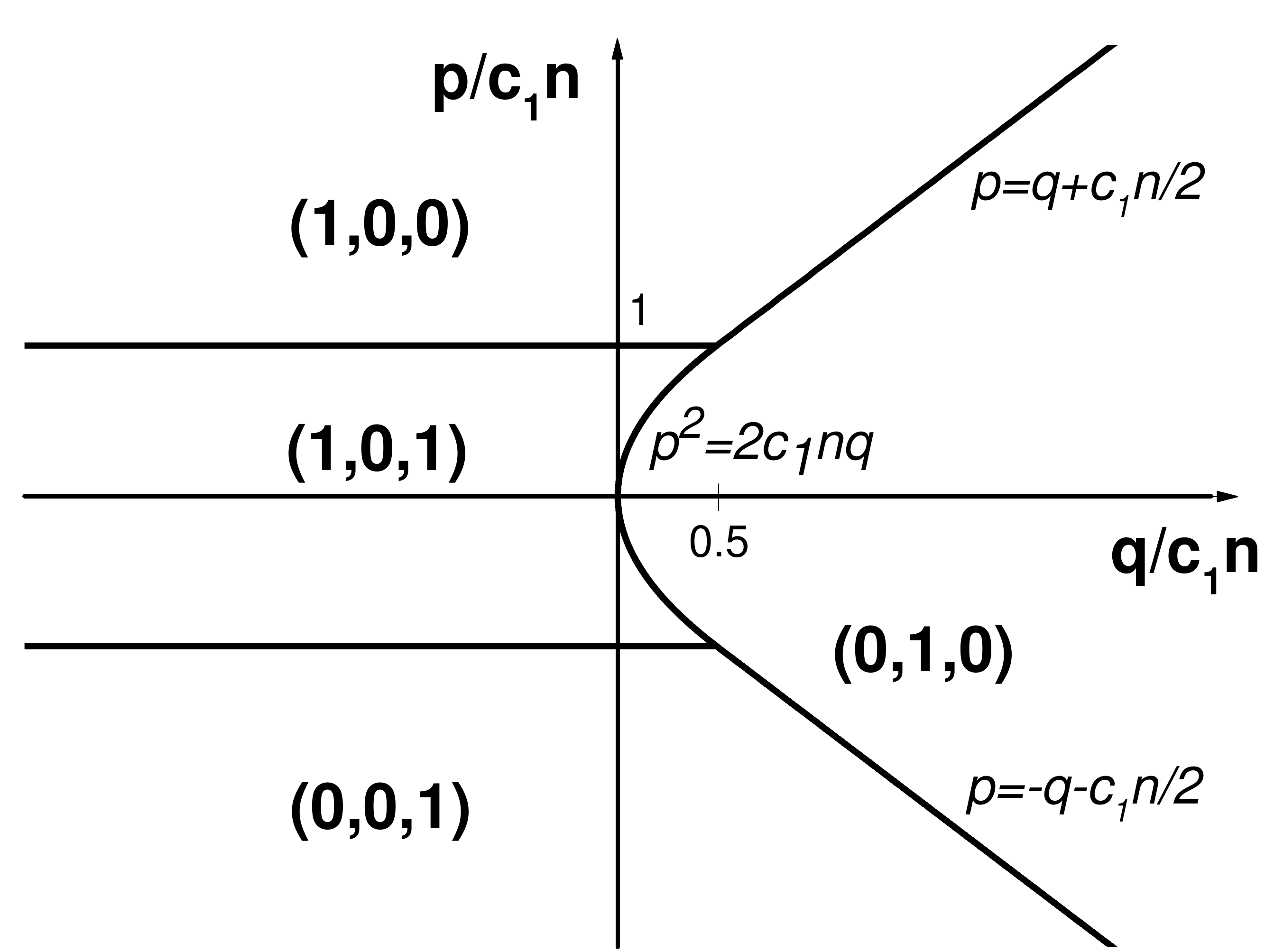}
		}
		\subfloat[$c_1=0$\label{subfig-2:hom2}]{%
			\includegraphics[width=0.33\textwidth]{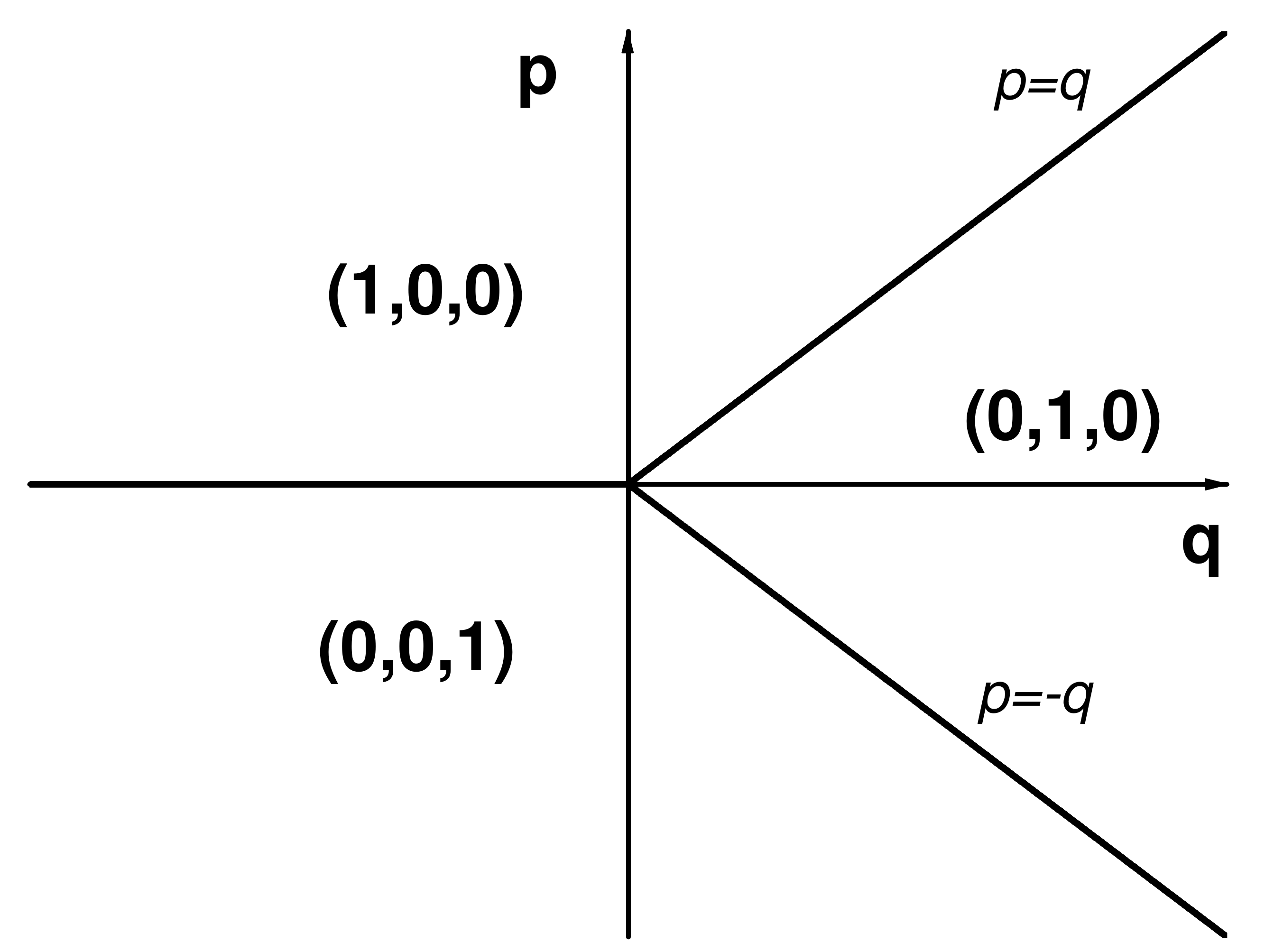}
		}
		\subfloat[$c_1<0$\label{subfig-3:hom3}]{%
			\includegraphics[width=0.33\textwidth]{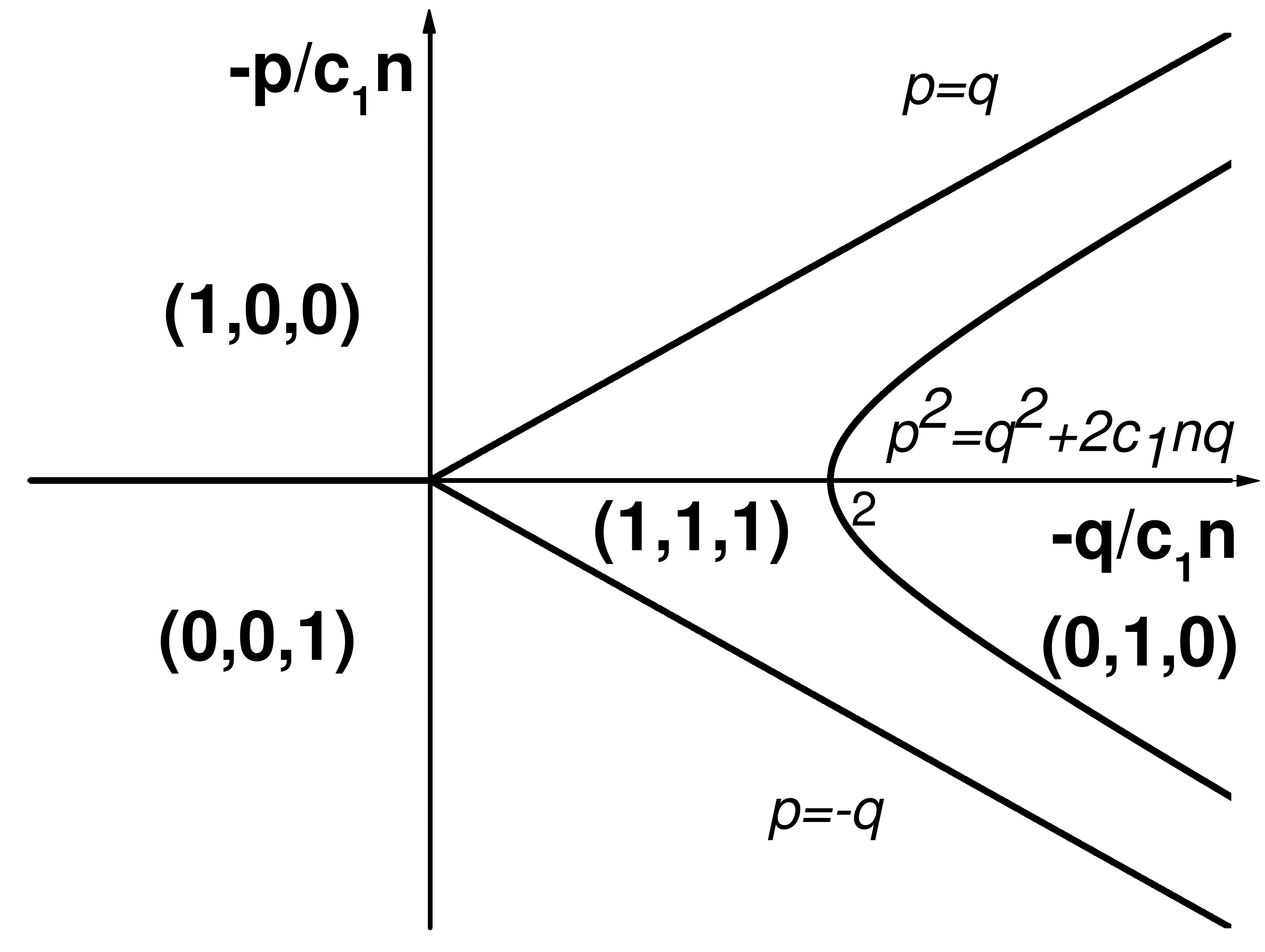}
		}
		\caption{Phase diagram in $(p,q)$ parametric space of a spin-1 BEC when the trapping potential is set to zero. }
		\label{fig:hom}
	\end{figure}
\end{widetext}

\section{3. Phase separation of the trapped condensate}\label{section:pstc}
In the presence of trapping potential $U(\vec{r})$ the GP dynamics of the spinor gas of spin-1 can be decomposed into parts by taking the ansatz,
\begin{equation}\label{eq:rev9}
\psi_m(\vec{r},t)=\sqrt{n_m(\vec{r})}exp(-\dfrac{i\mu t}{\hbar})exp(-i\theta_m).
\end{equation}
\par

The relative phase being defined as $\theta_r=\theta_1+\theta_{-1}-2\theta_0$, the dynamics of amplitudes and phases are,
\begin{equation}\label{eq:rev10}
\dot{n}_0(\vec{r})=-\dfrac{4c_1 n_0 \sqrt{n_1n_{-1}}\sin\theta_r}{\hbar},
\end{equation}
\begin{equation}\label{eq:rev11}
\dot{n}_{\pm1}(\vec{r})=\dfrac{2c_1 n_0 \sqrt{n_1n_{-1}}\sin\theta_r}{\hbar},
\end{equation}
\begin{equation}\label{eq:rev12}
\begin{split}
\hbar\dot{\theta}_0=\dfrac{1}{\sqrt{n_0(\vec{r})}}&\left(\mathcal{H}-\mu\right)\sqrt{n_0(\vec{r})}\\
&+c_1\left(n_1+n_{-1}+2 \sqrt{n_{-1}n_1}\cos\theta_r\right),
\end{split}
\end{equation}
\begin{equation}\label{eq:rev13}
\begin{split}
\hbar\dot{\theta}_{\pm1}=\dfrac{1}{\sqrt{n_{\pm1}(\vec{r})}}&\left(\mathcal{H}-\mu\right)\sqrt{n_{\pm1}(\vec{r})}\pm c_1\left(n_1-n_{-1}\right)+q\\
&\quad \mp p+c_1n_0\left(1 +\sqrt{\dfrac{n_{\mp1}(\vec{r})}{n_{\pm1}(\vec{r})}}\cos\theta_r\right);
\end{split}
\end{equation}
where $\mathcal{H}=-\dfrac{\hbar^2 \nabla^2}{2M} + U(\vec{r})+c_0 n$. 
The phase matching condition demands $\mu_{+}+\mu_{-}=2\mu_0$, which is valid even when $\mu_{+}=\mu_{-}=\mu_0$. $\mu$'s are the corresponding chemical potential and stability of the mixed phases would require it to remain constant. In what follows, we will always impose this condition of the constant chemical potential $\mu$ in order to have chemical stability of the co-existing phases. The relative phase $\theta_r$ and individual phases $\theta_m$'s are global parameters which actually hold the key of the relative energy of the various spin phases, that we are going to look for, on an equal footing. This consideration of not taking into account the space dependence of the phases, is quite consistent with the Thomas-Fermi limit, because after all, we will also be neglecting the derivatives of amplitudes considering the variation to remain slow.
\begin{widetext}
	\begin{table}[h!]
		{\begin{tabular}{ |p{1.5cm}|p{4.5cm}|p{9.5cm}|p{2cm}| }
				\hline
				\multicolumn{4}{|c|}{Stationary states for $c_1=0$} \\
				\hline
				States & Variation of density & Energy density & Restriction \\
				\hline
				(1,0,0) \quad F1 &$c_0n(\vec{r})=\mu+p-q-U(\vec{r})$  &$e_1=\dfrac{\left[\mu+p-q-U(\vec{r})\right]^2}{2c_0}+\dfrac{\left[ U(\vec{r})-p+q\right]\left[\mu+p-q-U(\vec{r})\right]}{c_0}$ & $none$ \\
				\hline
				(0,1,0) \quad P &$c_0n(\vec{r})=\mu-U(\vec{r})$   &$e_2=\dfrac{ U(\vec{r})\left[\mu-U(\vec{r})\right]}{c_0}+\dfrac{\left[\mu-U(\vec{r})\right]^2}{2c_0}$ &$none$ \\
				\hline
				(0,0,1) \quad F2 &$c_0n(\vec{r})=\mu-p-q-U(\vec{r})$ &$e_3=\dfrac{\left[\mu-p-q-U(\vec{r})\right]^2}{2c_0}+\dfrac{ \left[ U(\vec{r})+p+q\right]\left[\mu-p-q-U(\vec{r})\right]}{c_0}$ &$none$ \\
				\hline
				(1,1,0)   &$c_0n(\vec{r})=\mu-U(\vec{r})$  & $e_4=\dfrac{ U(\vec{r})\left[\mu-U(\vec{r})\right]}{c_0}+\dfrac{\left[\mu-U(\vec{r})\right]^2}{2c_0}$ &$p=q$ \\
				\hline
				(1,0,1)  & $c_0n(\vec{r})=\mu-q-U(\vec{r})$ & $e_5=\dfrac{ \left[ U(\vec{r})+q\right]\left[\mu-q-U(\vec{r})\right]}{c_0}+\dfrac{\left[\mu-q-U(\vec{r})\right]^2}{2c_0}$ &$p=0$ \\
				\hline
				(0,1,1) &$c_0n(\vec{r})=\mu-U(\vec{r})$ &$e_6=\dfrac{ U(\vec{r})\left[\mu-U(\vec{r})\right]}{c_0}+\dfrac{\left[\mu-U(\vec{r})\right]^2}{2c_0}$   &$p=-q$ \\
				\hline
				(1,1,1) & $c_0n(\vec{r})=\mu-U(\vec{r})$ &$e_7=\dfrac{ U(\vec{r})\left[\mu-U(\vec{r})\right]}{c_0}+\dfrac{\left[\mu-U(\vec{r})\right]^2}{2c_0}$ &$p=q=0$ \\
				\hline
			\end{tabular}
		}
		\caption{Stationary states at $c_1=0$. Associated condition for the last four states are $p=q$, $p=0$, $p=-q$ and $p=0,q=0$ respectively.}
	\end{table}
\end{widetext}

\subsection{3.1 Phase separation for $c_1=0$}\label{c0}
The condition, $c_1\simeq 0$ incorporates almost no interaction of spins. This is the situation sitting at the boundary of the two broad regimes namely $c_1>0$ (anti-ferromagnetic) and $c_1<0$ (ferromagnetic). We follow here the standard scheme of dividing the parameter regime of spin interactions as is done for the free condensate \cite{KAWAGUCHI2012253} to have a direct comparison. Setting $c_1=0$, one can now easily get the corresponding energy densities of the seven basic spin configuration in terms of the total density under T-F approximation (i.e. spatial derivatives of density and phases are neglected).  As an example let's explore the anti-ferromagnetic state, $(1,0,1)$. As $n_0=0$ here, Eq~\ref{eq:rev12} is no longer valid and the solution should obey the stationarity of other two sub-component phases resulting in
\begin{equation}\label{eq:rev14}
	U(\vec{r})+c_0n-\mu \mp p+q=0,
\end{equation} 
 when $n=n_1+n_{-1}$. Note that, Eqs~\ref{eq:rev13} take such a simple form because we are studying the case where spin dependent interaction is absent. From Eq~\ref{eq:rev14} it is easy to see that the T-F profile for the $(1,0,1)$ state would be,
\begin{equation}\label{eq:rev15}
	c_0n(\vec{r})=\mu-q-U(\vec{r}),
\end{equation}
 when $p=0$. Here $p=0$ is the condition for existence of this phase. Following the similar scheme would allow one to find the T-F profile, corresponding energy density and the parameter restriction for all the stationary states summarized in Table I.

 Note that, all the restrictions present on the parameters corresponding to the last four phases in the table which are $p=q$, $p=0$, $p=-q$ and $p=q=0$ arise from the solution of Eqs~\ref{eq:rev12}-\ref{eq:rev13}. 
An immediate consequence of these parameter restrictions is that, except for the case i.e. $p=q=0$, the states $(1,1,0)$, $(1,0,1)$ and $(0,1,1)$ cannot exist together. So there is no domain formation for these phases anywhere over the $p \ vs \ q$ parameter plane except at the origin. The phase $(1,1,1)$ exists only at the origin on this plane as well.

	\begin{figure}[h!]
		\subfloat[$ $\label{subfig-1:possibility}]{%
			\includegraphics[width=0.48\textwidth]{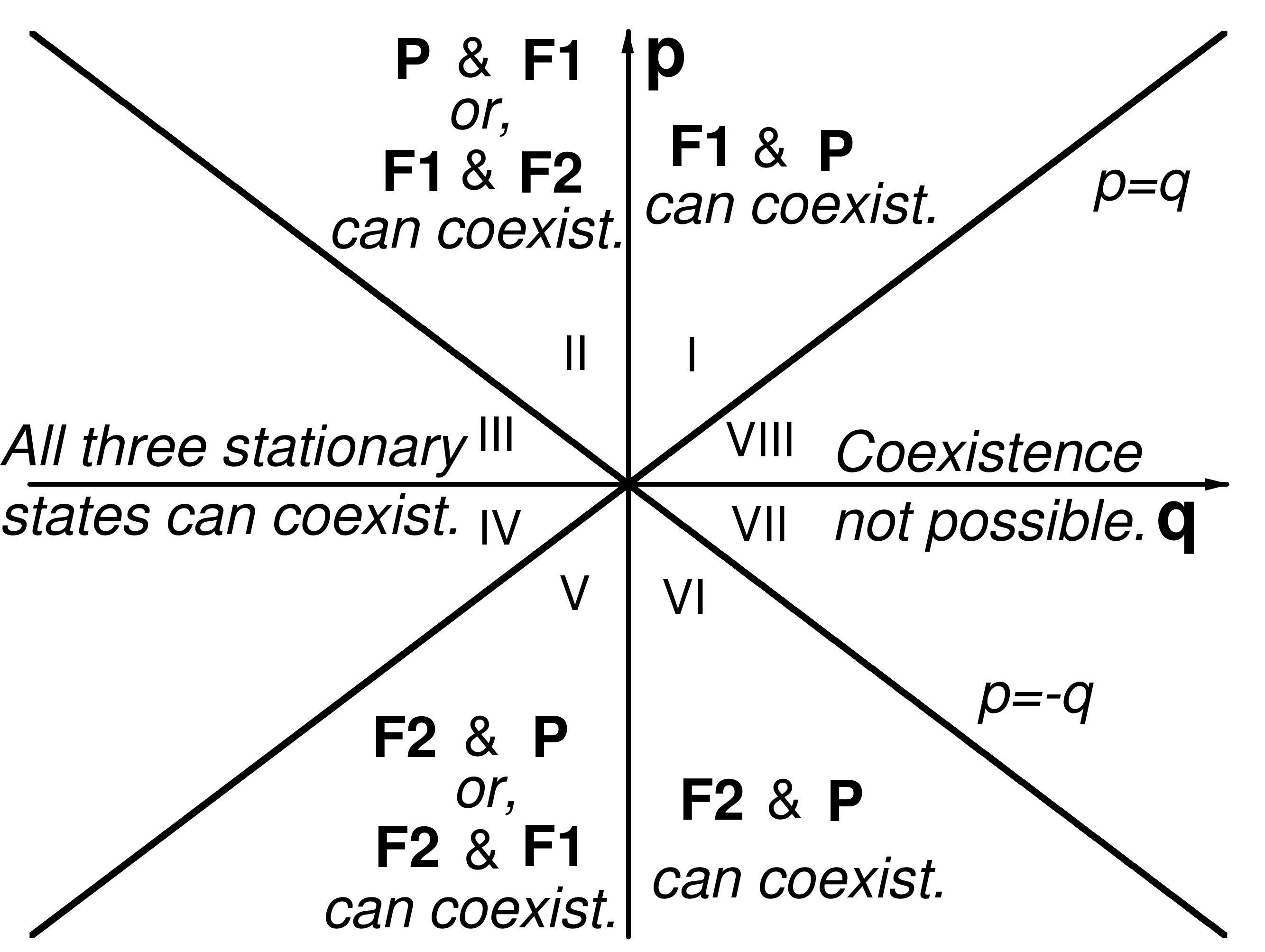}
		}
	
		\subfloat[$ $\label{subfig-2:circular}]{%
			\includegraphics[width=0.48\textwidth]{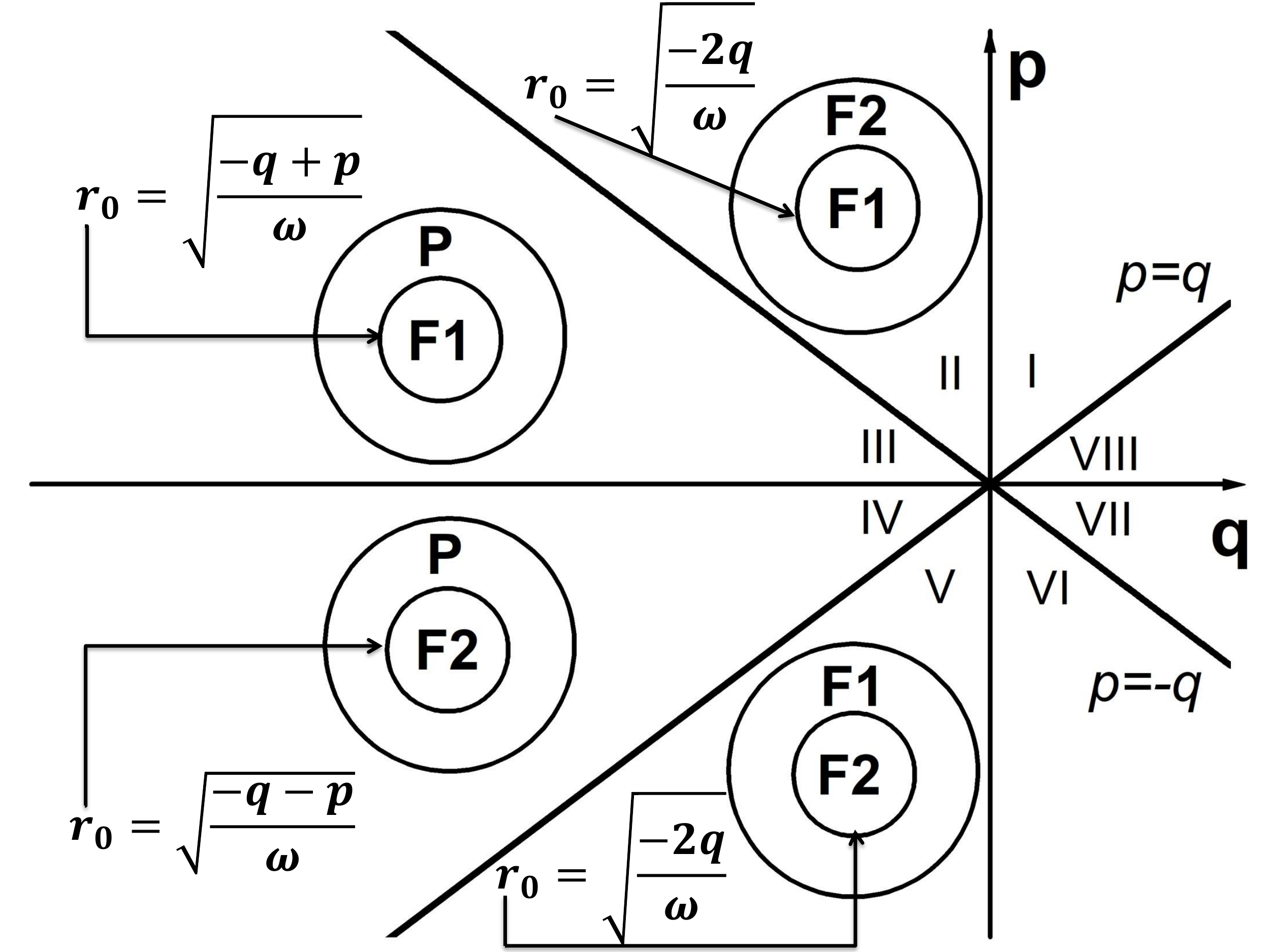}
		}
		
		\caption{Co-existing phases and domain formation in $(p,q)$ parameter space of trapped spin-1 BEC for $c_1=0$. The states $(1,0,0)$, $(0,0,1)$ and $(0,1,0)$ are represented by $F1$, $F2$ and $P$ for better visibility. For a uniform two dimensional harmonic trap the phase-separation radius ($r_0$) from the centre of the trap is shown as well.}
		\label{fig:possible}
    \end{figure}  

Fig~\ref{fig:possible}\subref{subfig-1:possibility} is a phase diagram showing which one of the first three phases $(1,0,0)$, $(0,1,0)$ and $(0,0,1)$ exists where on the  
$p \ vs \ q$ plane at $c_1=0$. This gives us a clear idea as to where on this phase diagram the domain formation can be expected, depending upon any particular form of the trapping potential $U(\vec{r})$, which is considered to be harmonic here. The diagonal lines, the $q-axis$ and the origin are the places where the other four phases namely $(1,1,0)$, $(1,0,1)$, $(0,1,1)$ and $(1,1,1)$
exist. An actual pair-wise comparison of energy densities shows the inner and the outer phases to expect in a harmonic trap in the region for negative $q$ in Fig~\ref{fig:possible}\subref{subfig-2:circular}. This figure also includes an estimation of the radius of phase boundaries under harmonic confinement. The same comparison is sufficient to deduce that no phase separation or domain formation is possible for $q>0$.

\begin{figure}[h!]
		\subfloat[$c_1=0$\label{subfig-1:p20c0}]{%
			\includegraphics[width=0.24\textwidth]{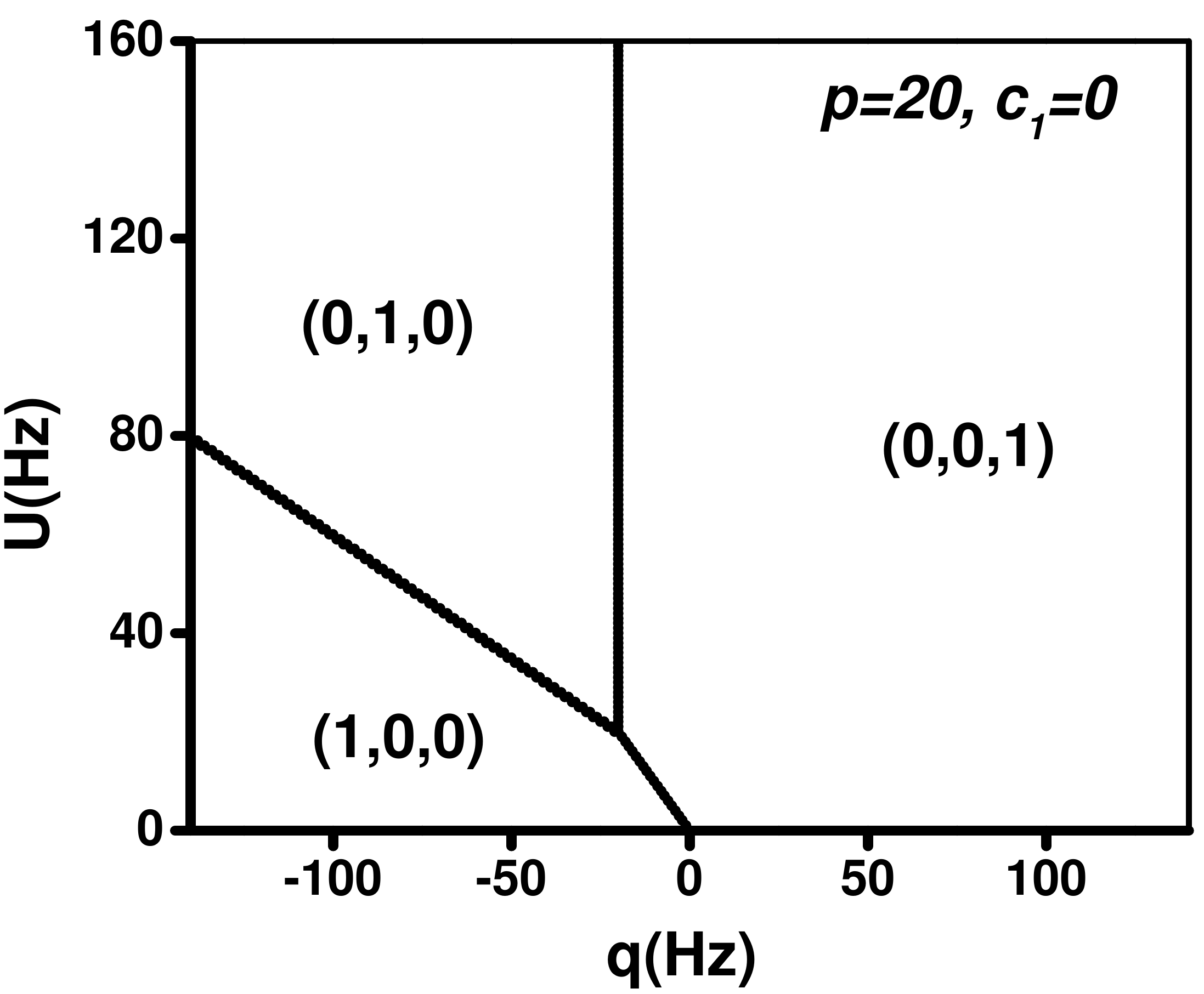}
		}
		\subfloat[$c_1=0$\label{subfig-2:pm20c0}]{%
			\includegraphics[width=0.24\textwidth]{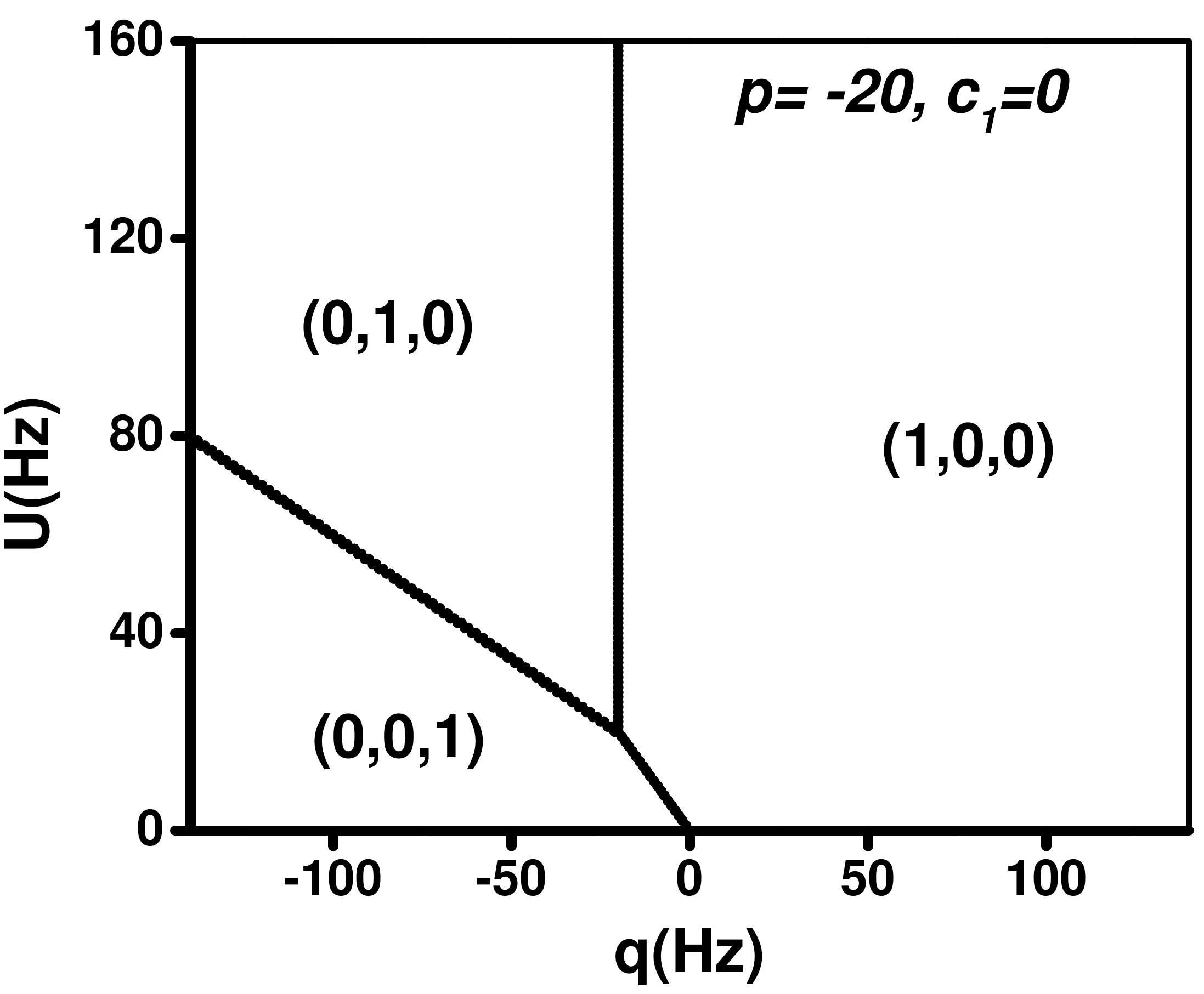}
		}
	
	    \centering
		\subfloat[$c_1=0$\label{subfig-3:qm20c0}]{%
			\includegraphics[width=0.24\textwidth]{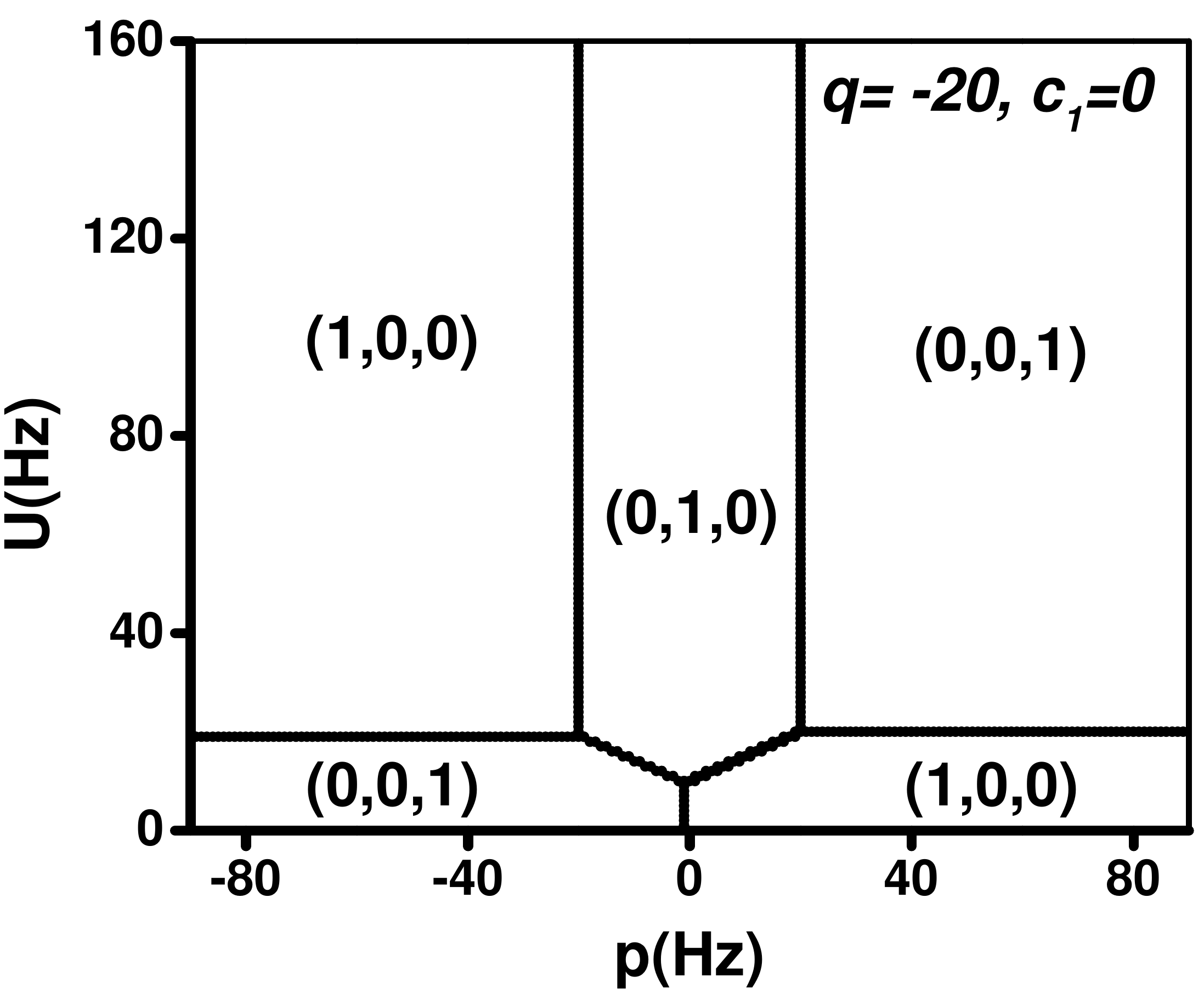}
		}
		\caption{Phase separation of a spin-1 BEC with almost no spin dependent interaction ($c_1=0$). $(a)$ and $(b)$ showing phase separation for opposite linear Zeeman terms. These sub-figures are symmetric under the change of the direction of magnetic field. ($c$) Same symmetry is revealed when quadratic Zeeman term is fixed at a negative value. The phase boundaries remain unaffected with the change of chemical potential ($\mu$). }
		\label{fig:c0}
	\end{figure}

To look at an example for the formation of domains when $U(\vec{r})=\frac{1}{2}\omega r^2$ is a harmonic trapping, let us choose a region (III) where first three single component phases can exist. A comparison of energy densities of the $(1,0,0)$ and $(0,1,0)$ shows that,
\begin{equation}\label{eq:rev16}
\Delta e_{12} \equiv e_1-e_2=\dfrac{(p-q)[2U(\vec{r})-(p-q)]}{2c_0}
\end{equation} 
and it implies that the state $(1,0,0)$ energetically is favored below a radius $r_0^2=(p-q)/\omega$ because $(p-q)>0$. The state $(0,1,0)$ should be existing for $r>r_0$ and is the peripheral state when $(1,0,0)$ sits at the core of the harmonic trap. The situation does not happen when $(p-q)$ is negative as the radius becomes imaginary. The same reason is enough to understand that phase separation between the $F1$ and $P$ is only possible in region-$I$,$II$,$III$ and $IV$ of Fig~\ref{fig:possible}\subref{subfig-1:possibility}. All such comparison can now be done and one gets the ground state domains of stationary phases under T-F approximations for $c_1=0$. Though in the regions marked as $III$ and $IV$ in Fig~\ref{fig:possible}\subref{subfig-1:possibility} all three types of phase separation is allowed, no cases can be found for simultaneous domain formation of all three states. One can start the analysis by first considering which of the states is energetically favored at the centre of the trap, as the condensation in an experimental situation arises first at the central region because of the density being maximum there \cite{proukakis2013quantum}. For example in region-$III$ where $(1,0,0)$ sits at the centre of the trap, out of two possibilities of separation $(0,1,0)$ wins because the separation can happen at a smaller radius than that with $(0,0,1)$. Now, when $(0,1,0)$ is in the outer region one can check that $(0,0,1)$ never wins energetically over $(0,1,0)$. To understand these, one can have a look at the phase diagrams (Fig~\ref{fig:c0}) on an $U(r)$ vs $p$ and $U(r)$ vs $q$ planes. Phase separation between ferromagnetic and polar phases are observed here as one moves upward along the $U(\vec{r})$-axis at relatively larger negative q values. At a smaller $-ve$ q value, two ferromagnetic phases form domains.

Note that no possible phase separation can happen in the region marked $VII$ and $VIII$. In this region the ground state will be selected depending on the chemical potential $\mu$. As we are only concentrating on the phase separation scenario keeping a constant $\mu$ for the three unrestricted stationary states, we find $(0,0,1)$ to be energetically lowest in the region $I$,$VIII$ (see Fig~\ref{fig:c0}\subref{subfig-1:p20c0}). This situation might change when the constant $\mu$ condition is relaxed in order to find the only existing phase without any phase separation. However, that is not of our interest in this paper.
\par
A quick comparison of this confined case can be done with the phase diagram Fig~\ref{fig:hom}\subref{subfig-2:hom2} of the uniform BEC. Fig~\ref{fig:hom}\subref{subfig-2:hom2} indicates a phase separation existing for positive $q$, whereas, T-F approximated calculations under actual confinement gives here results in contrary to that. Fig~\ref{fig:hom}\subref{subfig-2:hom2} also indicates that there can be no phase co-existence of the two opposite ferromagnetic phases (at nonzero $p$), however, the confined picture reveals the opposite. This is exactly the reason one should be guided by the phase separation scenario under actual confinement rather than extrapolating density dependence of the phase in the homogeneous case to the phase separation under confinement. 

\subsection{3.2 Phase co-existence for $c_1\neq 0$}
The condition $c_1\neq 0$ involves both anti-ferromagnetic and ferromagnetic interaction for $c_1>0$ and $c_1<0$ respectively. For nonzero spin interaction, it is obvious from Eq.\ref{eq:rev10}-\ref{eq:rev11} that the temporal variation of the different spin densities should go to zero for the stationary states. So one is left with two choices, 
\begin{itemize}
	\item at least one of the spin density is zero (corresponds to the first six states in Table-II) or,
	\item all the subcomponents are populated but the relative phase is either $0$ or $\pi$. 
\end{itemize} 

Let's see how in details the second option with $\theta_r=0$ because this situation is the most complex one. For this case equations corresponding to all the phases are valid as all the subcomponents are all populated. 
By exploiting the stationarity of phases one gets the corresponding equation for the $n_0$ subcomponent,
\begin{equation}\label{eq:rev17}
\left[U_{t}(\vec{r})+c_0n-\mu\right]+c_1\left(n_1+n_{-1}+2 \sqrt{n_{-1}n_1}\cos\theta_r\right)=0.
\end{equation}
For further simplification one can define a parameter, $k(\vec{r})=\sqrt{\dfrac{n_1}{n_{-1}}}$. Note that the ansatz (Eq~\ref{eq:rev9}) allows $\sqrt{n_m(\vec{r})}$ to take only positive values, negative value being accounted for by the phase factor. So by definition $k(\vec{r})$ is positive and nonzero here. The condition $\theta_r=0$ leads to,
\begin{equation}\label{eq:rev18}
\left[U_{trap}(\vec{r})+c_0n(\vec{r})-\mu\right]=-c_1\left[k(\vec{r})+1\right]^2n_{-1}(\vec{r}).
\end{equation}
Now the other two phase equations~(\ref{eq:rev13}) become,
\begin{equation}\label{eq:rev19}
\begin{split}
U_t(\vec{r})-p+q+c_0n(\vec{r})&+c_1(k^2(\vec{r})-1))n_{-1}(\vec{r})-\mu\\
&+c_1n_0(\vec{r})+c_1n_0(\vec{r})\dfrac{1}{k}=0
\end{split}
\end{equation}
\begin{equation}\label{eq:rev20}
\begin{split}
U_t(\vec{r})+p+q+c_0n(\vec{r})&-c_1(k^2(\vec{r})-1))n_{-1}(\vec{r})-\mu\\
&+c_1n_0(\vec{r})+c_1n_0(\vec{r})k=0.
\end{split}
\end{equation}
Subtracting Eq~\ref{eq:rev19} from Eq~\ref{eq:rev20} one can express $n_0$ in terms of $k$ and $n_{-1}$,
\begin{equation}\label{eq:rev21}
c_1n_0(\vec{r})=\dfrac{-2p+2c_1(k^2-1)n_{-1}(\vec{r})}{k-\dfrac{1}{k}}.
\end{equation} 
Similarly addition leads to another expression of $n_0$,
\begin{equation}\label{eq:rev22}
c_1n_0(\vec{r})=\dfrac{2c_1(k+1)^2n_{-1}(\vec{r})-2q}{k+\dfrac{1}{k}+2}.
\end{equation}
Solving last two equations one gets to express $k(\vec{r})$ in terms of the external parameters $p$ and $q$ as, $k=\dfrac{q+p}{q-p}$. It is easy to see that $k$ is positive only for $|q|>|p|$.\\
So, replacing the value of this $k$ in any of the equations of $n_0$, and then using the equation, $n=n_0+(k^2+1)n_{-1}$ we get the number densities to be,
\begin{equation}\label{eq:rev23}
n_1(\vec{r})=\dfrac{(p+q)^2}{4q^2}\left[n(\vec{r})+\dfrac{q^2-p^2}{2c_1q}\right]
\end{equation}
\begin{equation}\label{eq:rev24}
n_{-1}(\vec{r})=\dfrac{(p-q)^2}{4q^2}\left[n(\vec{r})+\dfrac{q^2-p^2}{2c_1q}\right]
\end{equation}
\begin{equation}\label{eq:rev25}
n_0(\vec{r})=\dfrac{(q^2-p^2)}{2q^2}\left[n(\vec{r})-\dfrac{q^2+p^2}{2c_1q}\right]
\end{equation}
This state corresponding to $\theta_r=0$ is valid for the condition $|q|>|p|$ as reasoned earlier. 
\par
The total number density(defined as $n(\vec{r})=n_1(\vec{r})+n_0(\vec{r})+n_{-1}(\vec{r})$) varies as,
\begin{equation}\label{eq:rev26}
n(\vec{r})=\dfrac{\mu-U_t(\vec{r})+\dfrac{(p^2-q^2)}{2q}}{(c_0+c_1)}.
\end{equation} 
\par
Corresponding energy density can be calculated by using this expression in Eq~\ref{eq:rev7}
\begin{equation}\label{eq:rev27}
\begin{split}
e(\vec{r})=U_t(\vec{r})&\dfrac{[k_1-U_t(\vec{r})]}{(c_0+c_1)}+\dfrac{1}{2}c_0\left[\dfrac{k_1-U(\vec{r})}{c_0+c_1}\right]^2\\
&\quad+\dfrac{1}{2}c_1\left[\dfrac{k_1-U(\vec{r})}{c_0+c_1}+\dfrac{p^2-q^2}{2qc_1}\right]^2,
\end{split}
\end{equation}
where, $k_1=\mu+\dfrac{(p^2-q^2)}{2q}$.
\par
 The method we have used is sufficient to extract information about $APM$ state ($\theta_r=\pi$) as well. We find for $APM$ state, $k=\dfrac{q+p}{p-q}$. The fact that $k$ being positive as discussed earlier ensures that $|p|>|q|$.
Though this two conditions ($\theta_r=0$ or $\pi$) lead to the same density and energy density profile of $(1,1,1)$ state, the phase matched and anti-phase matched states exist only for the conditions $|p|<|q|$ and $|p|>|q|$ respectively. 

\begin{widetext}
	\begin{table}[h!]
		{\begin{tabular}{ |p{1.3cm}|p{4.5cm}|p{9cm}|p{2.6cm}| }
				\hline
				\multicolumn{4}{|c|}{Stationary states for $c_1\neq 0$} \\
				\hline
				States &Variation of density &Energy density &Restriction \\
				\hline
				(1,0,0)  $F1$ &$(c_0+c_1)n(\vec{r})=\mu+p-q-U(\vec{r})$  &$\dfrac{\left[ U(\vec{r})-p+q\right]\left[\mu+p-q-U(\vec{r})\right]}{(c_0+c_1)}+\dfrac{\left[\mu+p-q-U(\vec{r})\right]^2}{2(c_0+c_1)}$ &$none$ \\
				\hline
				(0,1,0)  $P$ &$c_0n(\vec{r})=\mu-U(\vec{r})$   &$\dfrac{ U(\vec{r})\left[\mu-U(\vec{r})\right]}{c_0}+\dfrac{\left[\mu-U(\vec{r})\right]^2}{2c_0}$ &$none$ \\
				\hline
				(0,0,1)  $F2$ &$(c_0+c_1)n(\vec{r})=\mu-p-q-U(\vec{r})$ &$\dfrac{ \left[ U(\vec{r})+p+q\right]\left[\mu-p-q-U(\vec{r})\right]}{(c_0+c_1)}+\dfrac{\left[\mu-p-q-U(\vec{r})\right]^2}{2(c_0+c_1)}$ &$none$\\
				\hline
				(1,1,0)  $MF1$  &$(c_0+c_1)n(\vec{r})=\mu-U(\vec{r})+(p-q)$  &$\dfrac{ U(\vec{r})\left[\mu+p-q-U(\vec{r})\right]}{(c_0+c_1)}+\dfrac{c_0\left[\mu+p-q-U(\vec{r})\right]^2}{2(c_0+c_1)^2}$ &  $n_0=\dfrac{p-q}{c_1}$ \\
				\hline
				(1,0,1)  $AF$ & $c_0n(\vec{r})=\mu-q-U(\vec{r})$ and \quad $(n_1-n_{-1})\equiv F_z=\dfrac{p}{c_1}$ & $\dfrac{ \left[ U(\vec{r})+q\right]\left[\mu-q-U(\vec{r})\right]}{c_0}+\dfrac{\left[\mu-q-U(\vec{r})\right]^2}{2c_0}-\dfrac{p^2}{2c_1} $ & $none$ \\
				\hline
				(0,1,1)  $MF2$ &$(c_0+c_1)n(\vec{r})=\mu-U(\vec{r})-(p+q)$  &$\dfrac{ U(\vec{r})\left[\mu-p-q-U(\vec{r})\right]}{(c_0+c_1)}+\dfrac{c_0\left[\mu-p-q-U(\vec{r})\right]^2}{2(c_0+c_1)^2}$ &$n_0=\dfrac{-p-q}{c_1}$   \\
				\hline
				(1,1,1)  $(A)PM$ & $(c_0+c_1)n(\vec{r})=k_1-U(\vec{r})$ where, $k_1=\mu+\dfrac{(p^2-q^2)}{2q}$ &$\dfrac{ U(\vec{r})\left[k_1-U(\vec{r})\right]}{c_0+c_1}+\dfrac{c_0}{2}\left[\dfrac{k_1-U(\vec{r})}{c_0+c_1}\right]^2+\dfrac{c_1}{2}\left[\dfrac{k_1-U(\vec{r})}{c_0+c_1}+\dfrac{p^2-q^2}{2qc_1}\right]^2 $ &$ $ $PM(|p|<|q|)$ $ $ $APM(|p|>|q|)$ \\
				\hline
			\end{tabular}
		}
		\caption{Different stationary states at $c_1\neq0$.}
		
	\end{table}
\end{widetext}

\subsubsection{\underline{3.2.1 Anti-ferromagnetic interaction $c_1> 0$}}\label{subsubsec:cgrtr}
 For anti-ferromagnetic type of interaction, energetic comparison of all the seven possible states reveals the phase separated ground state structure. Important to note that the mixed ferromagnetic-polar states $(1,1,0)$ and $(0,1,1)$ only exists for the $p$, $q$ values, for which $n_0$ is non-negative (see Table II). As is already mentioned, all the energy density comparison is done at a constant chemical potential ($\mu$), ensuring chemical stability. We fix the $\mu$ at 400 nK and investigate the case for $^{23}Na$, for which $c_1$ is positive ($2.415\times 10^{-19}$ Hz). The parameter $c_0$ is numerically $149.89 \times 10^{-19} Hz$ for this element. The controllable parameters $p$ and $q$ can be safely varied from $-150 Hz$ to $150 Hz$. External potential $U$ is varied from $0$ to $170Hz$. To observe the phase separation phenomenon we fix either $p$ or $q$ and tune the other with $U$.

	\begin{figure}[h!]
		\subfloat[$c_1>0 $\label{subfig-1:q30cg1}]{%
			\includegraphics[width=0.24\textwidth]{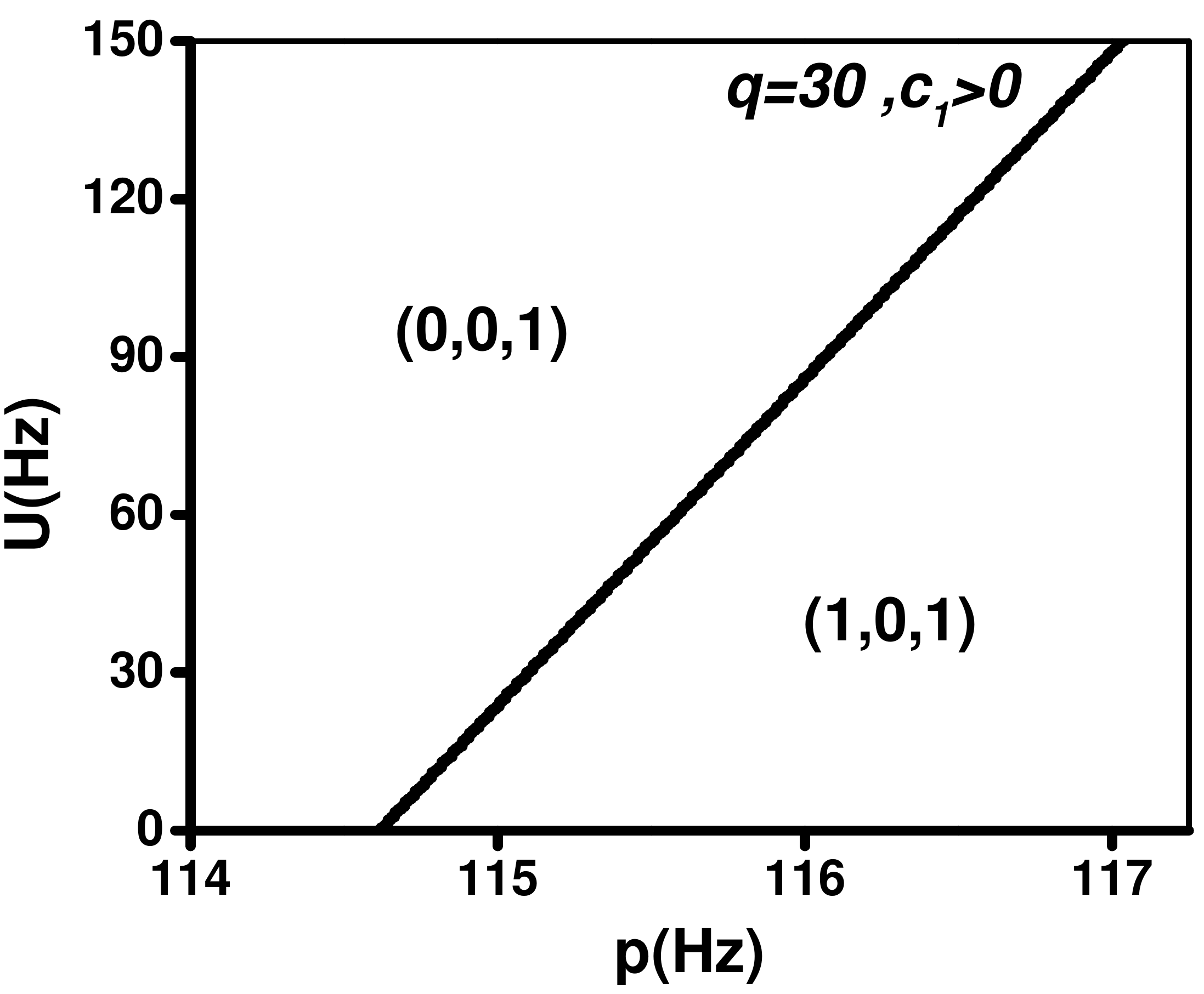}
		}
		\subfloat[$c_1>0 $\label{subfig-2:q30cg2}]{%
			\includegraphics[width=0.24\textwidth]{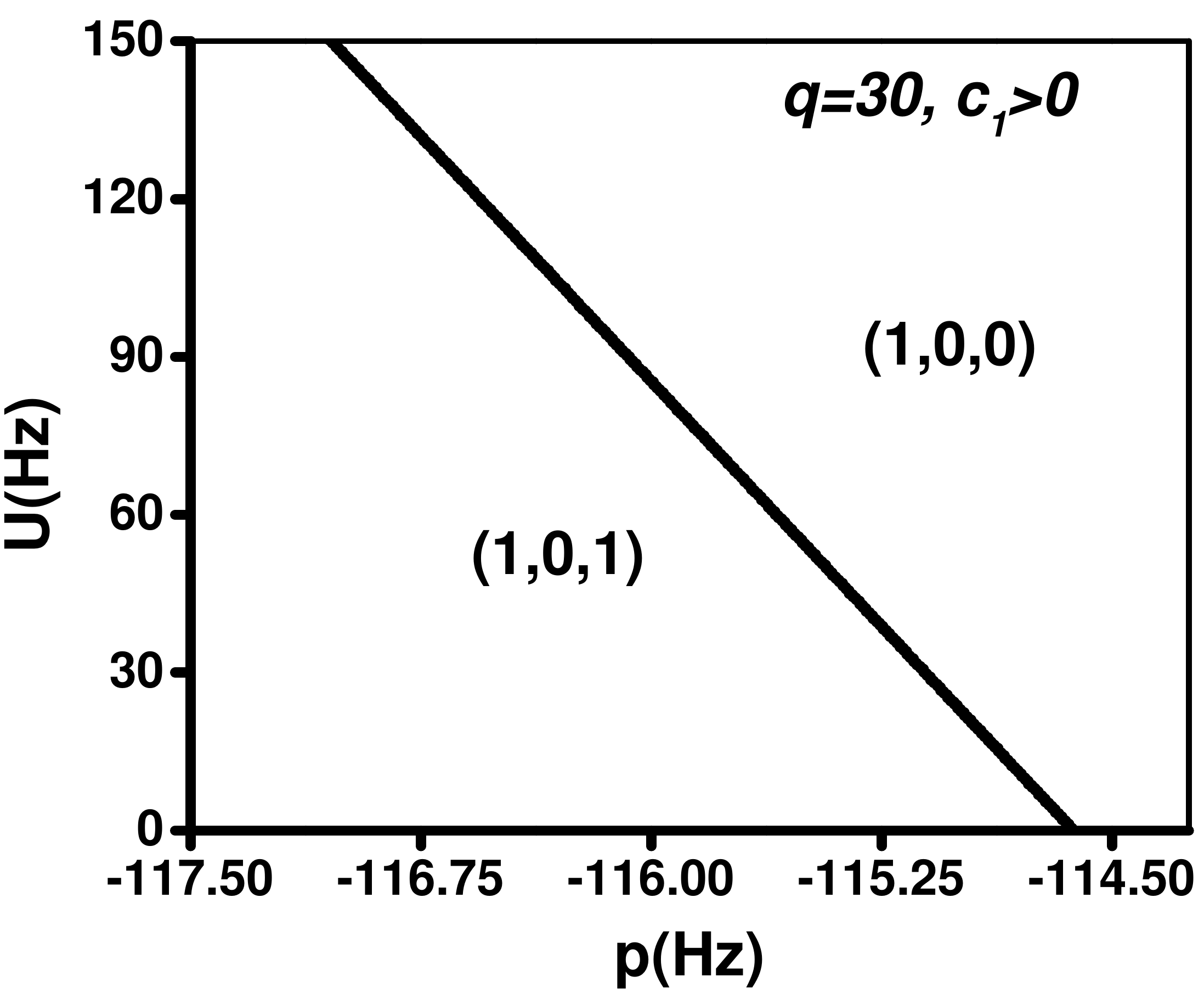}
		}
	    	
		\centering
		\subfloat[$c_1>0 $\label{subfig-3:q100cg1}]{%
			\includegraphics[width=0.24\textwidth]{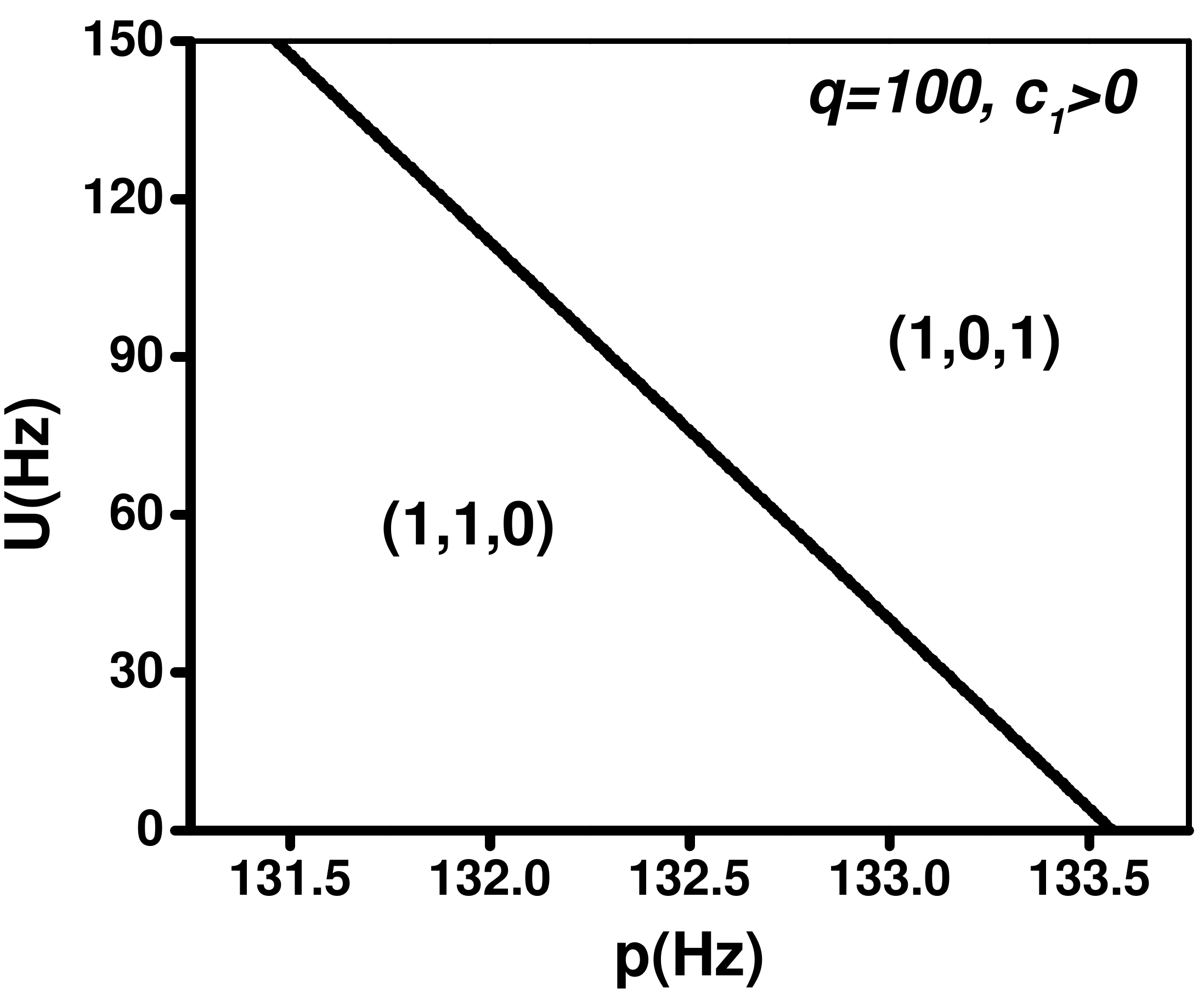}
		}
		
		\caption{Two state coexistence with one of them being the $AF$ state for anti-ferromagnetic type of interaction.}
		\label{fig:aferrocg}
	\end{figure} 

It is quite expected that the anti-ferromagnetic interaction will favour the formation of phase co-existence with anti-ferromagnetic ($AF$) state. Figs~\ref{fig:aferrocg}\subref{subfig-1:q30cg1}, \subref{subfig-2:q30cg2} show for high values of $|p|$ when $q$ is fixed at $30 Hz$, the $AF$-phase dominates at the centre of the trap while the ferromagnetic states $F2$ or $F1$ wins over all other states to form the periphery. Obviously, reversal of the sign of linear Zeeman term allows $F1$ to beat the $F2$ and vice versa, which is not at all unexpected. There is an interesting common feature here, with the increment of $|p|$ the $AF$ state expands its domain while ferromagnetic states shrink in both the cases. If $q$, the quadratic Zeeman term is tuned to a larger value, say at $100 Hz$, one can notice that a mixed ferromagnetic state ($MF1$) energetically beats the $AF$ phase to capture the centre spot (Fig~\ref{fig:aferrocg}\subref{subfig-3:q100cg1}). $AF$ state remains at the peripheral region. It should be noted that, the subcomponent number density should obey $n_0\geq 0$, which imposes restriction for both the $MF1$ and $MF2$ states. In this case both $c_1$ and the term $(p-q)$ are positive, allowing $MF1$ to appear. As the intuition suggests, reversal of the sign of $p$ (in a region from $-133.5$ to $-131.5 Hz$) does indeed prefer the other mixed ferromagnetic state $MF2$ in place of $MF1$ keeping the same structure as Fig~\ref{fig:aferrocg}\subref{subfig-3:q100cg1}.
 
 	\begin{figure}[h!]
 		\subfloat[$c_1>0 $\label{subfig-1:q30cg3}]{%
 			\includegraphics[width=0.24\textwidth]{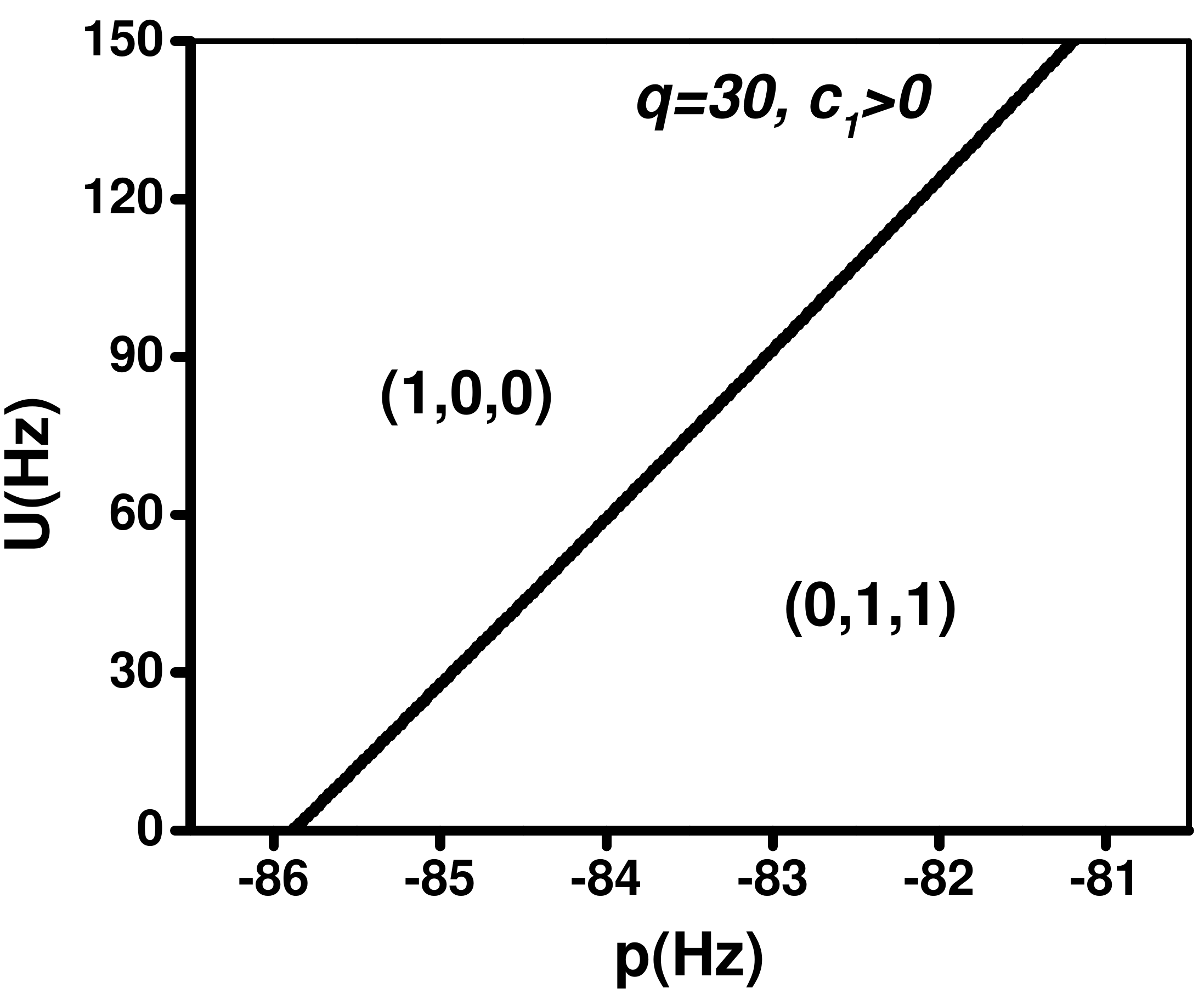}
 		}
 		\subfloat[$c_1>0 $\label{subfig-2:p100cg2}]{%
 			\includegraphics[width=0.24\textwidth]{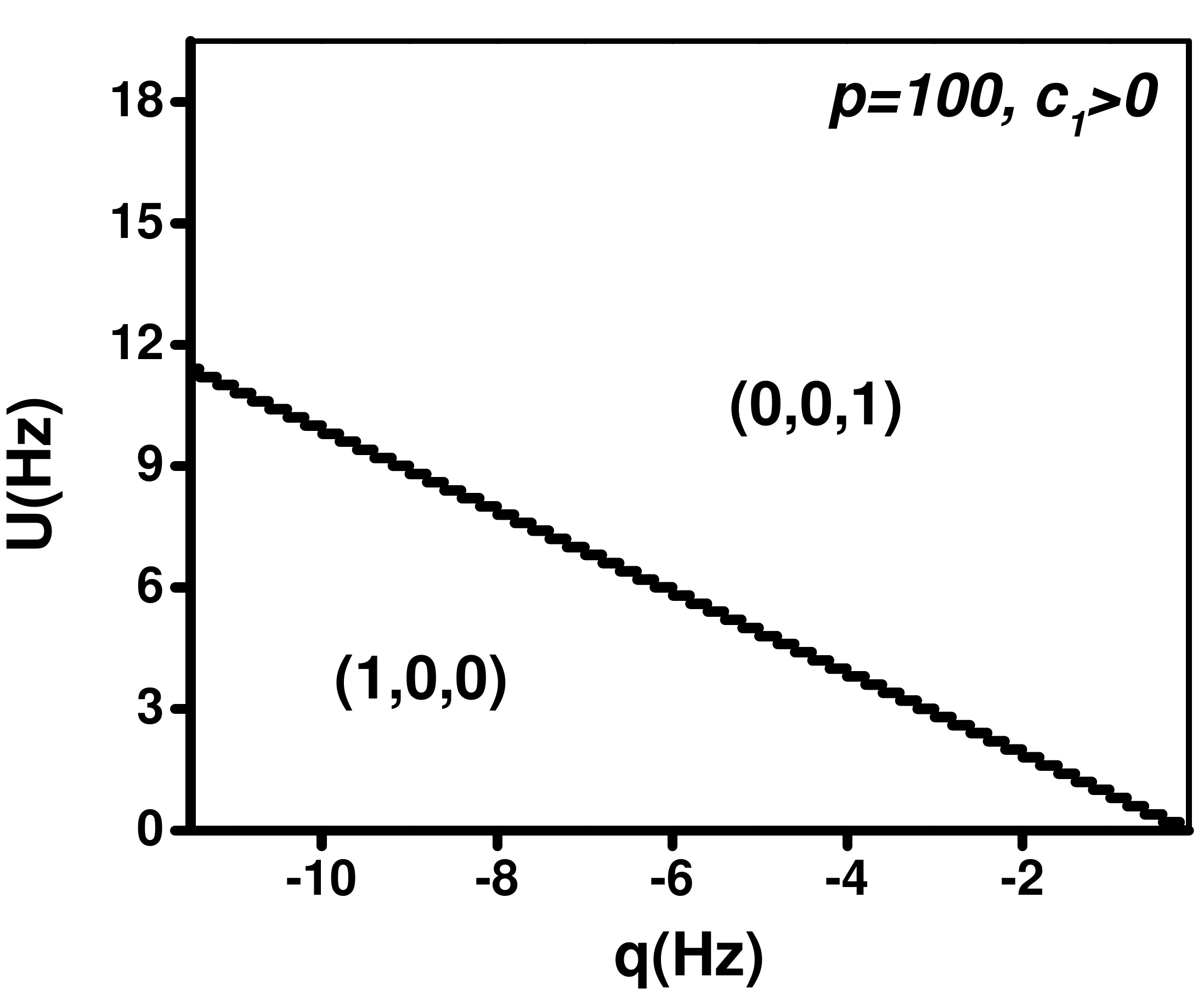}
 		}
 		
 		\subfloat[$c_1>0 $\label{subfig-3:p30cg1}]{%
 			\includegraphics[width=0.24\textwidth]{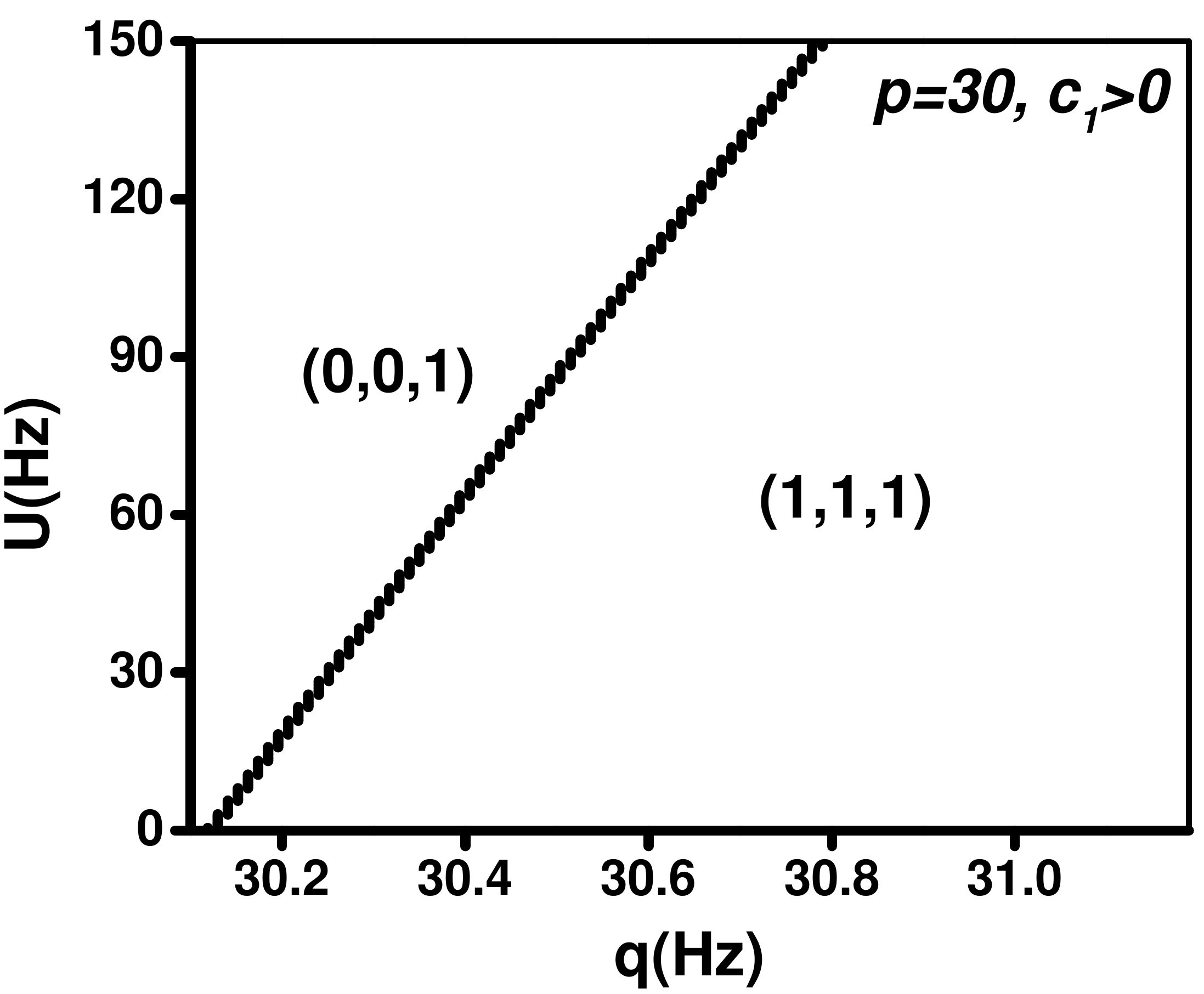}
 		}
 		\subfloat[$c_1>0 $\label{subfig-4:q100cg3}]{%
 			\includegraphics[width=0.24\textwidth]{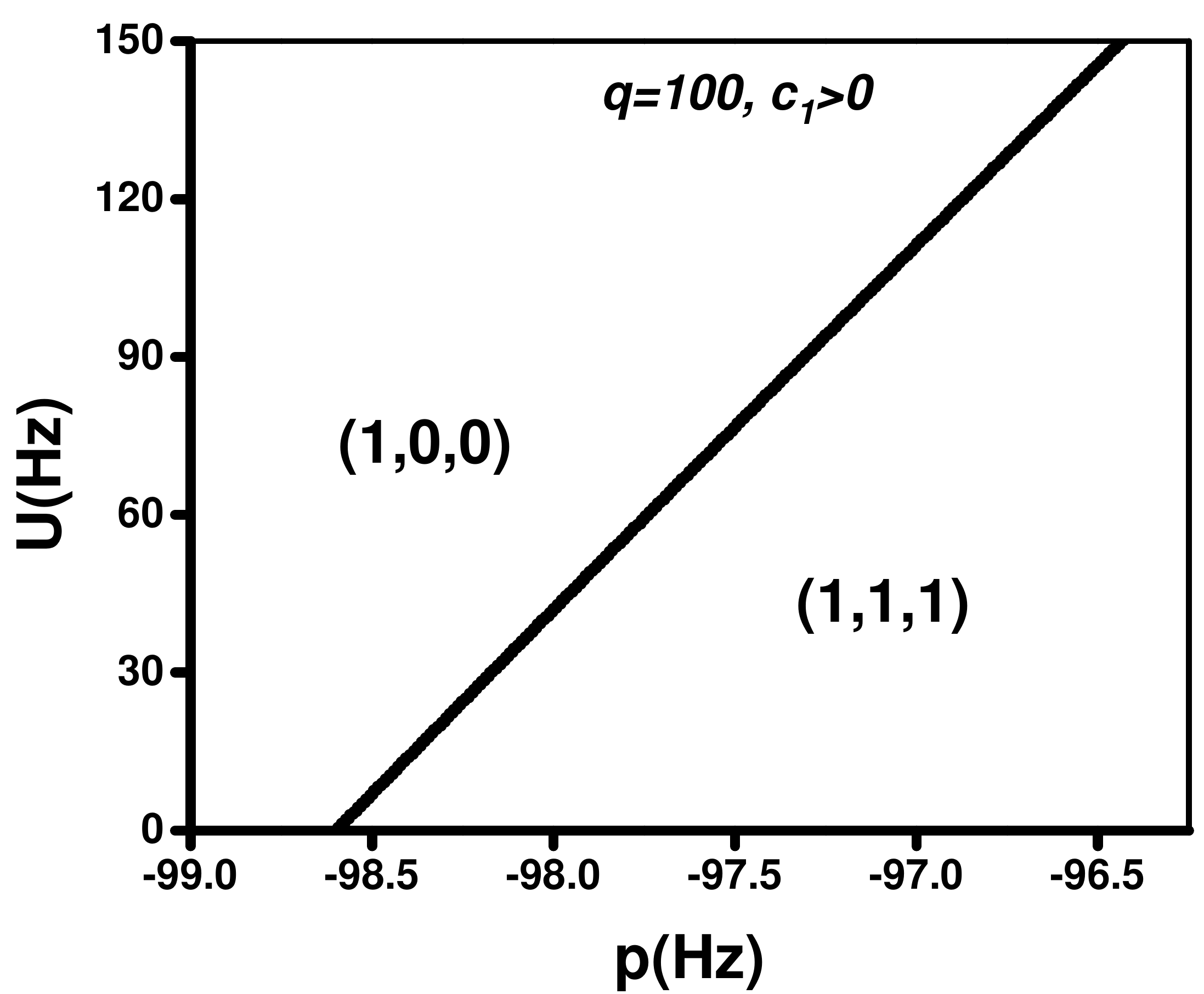}
 		}
 		
 		\subfloat[$c_1>0 $\label{subfig-5:p100cg1}]{%
 			\includegraphics[width=0.24\textwidth]{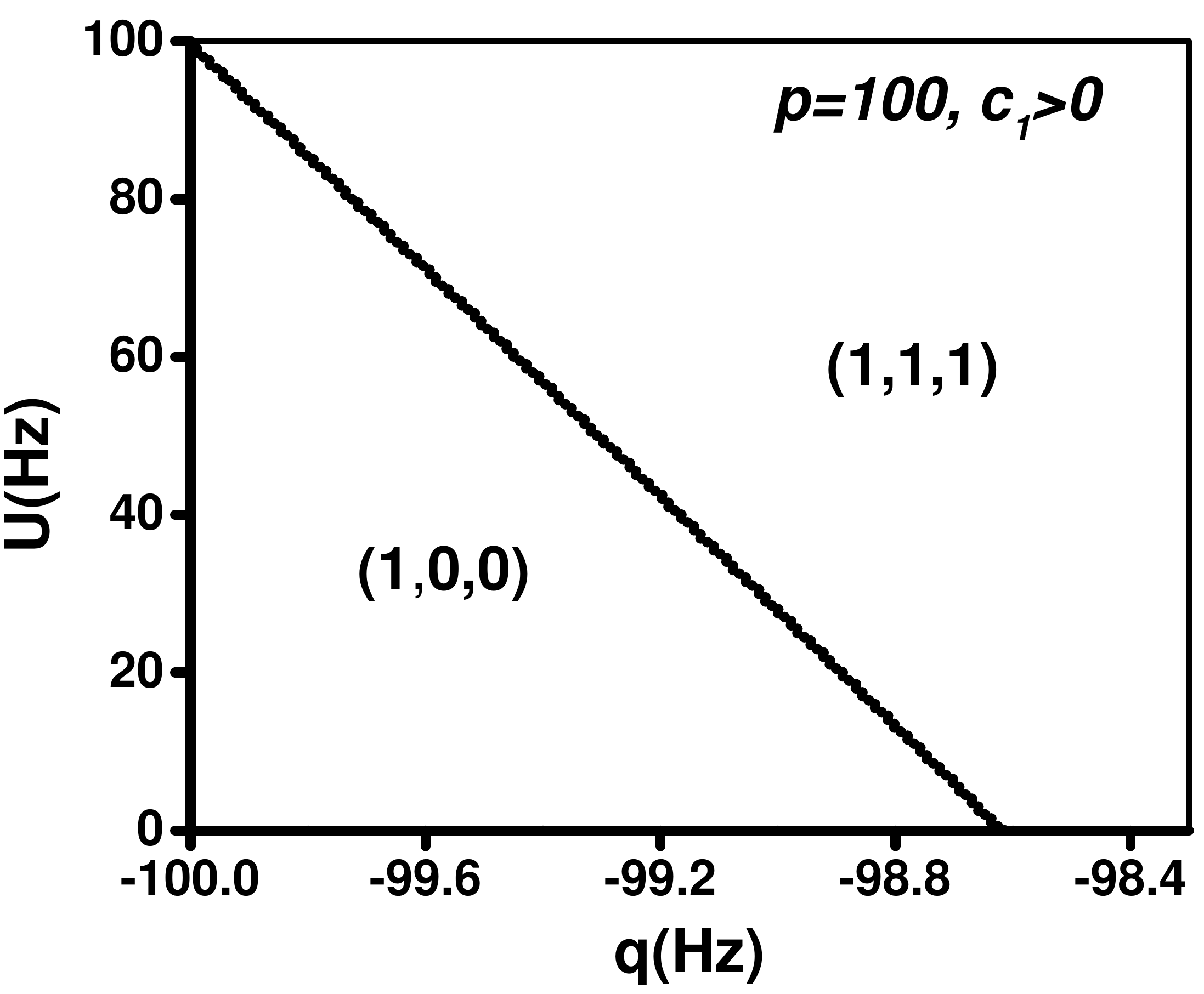}
 		}
 		\subfloat[$c_1>0 $\label{subfig-6:qm5cg1}]{%
 			\includegraphics[width=0.24\textwidth]{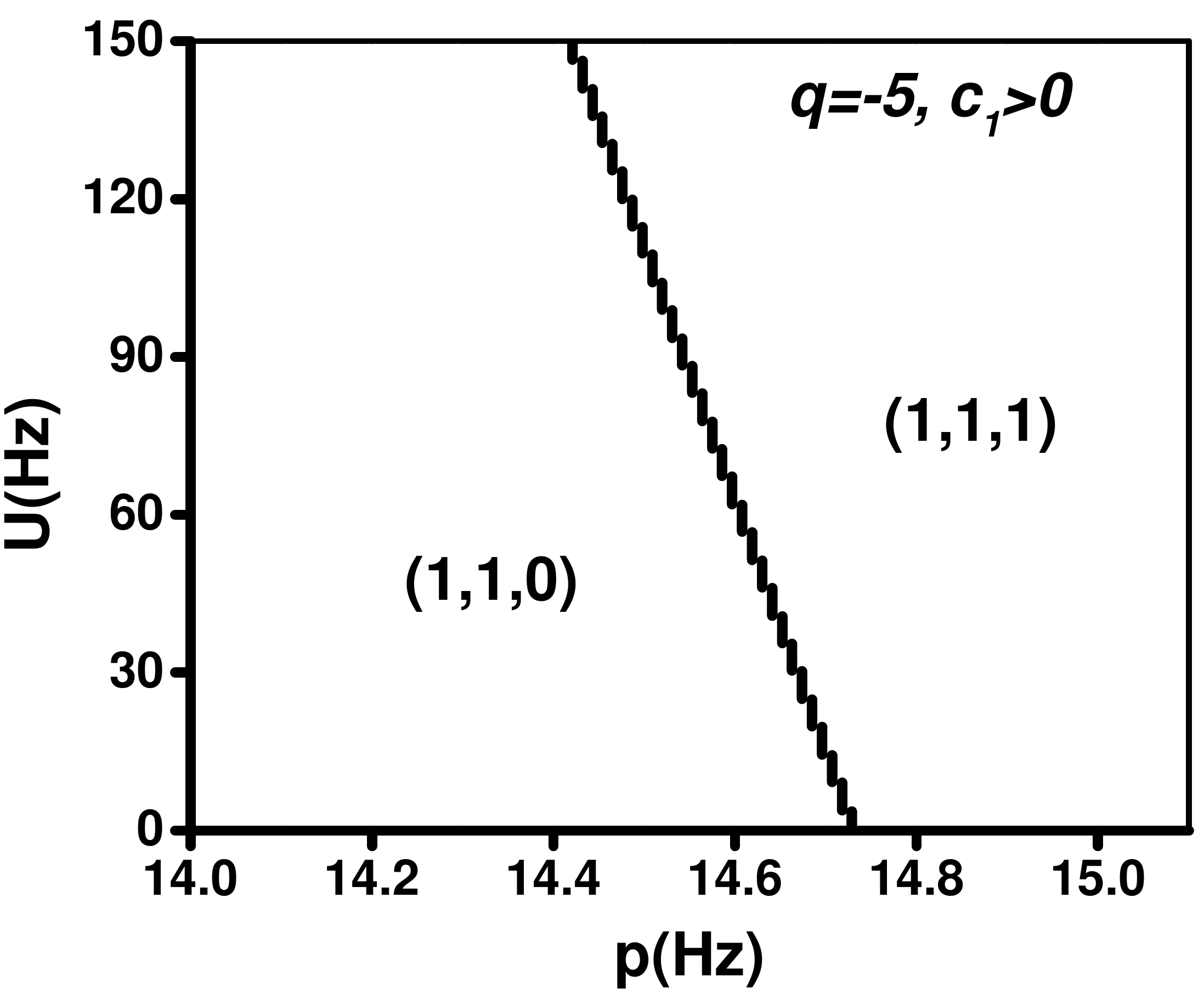}
 		}  
 		\caption{Possible two state coexisting structures; with (b),(c),(e) at fixed $p$ and (a),(d),(f) at fixed $q$. Though, the interaction type is anti-ferromagnetic these domain forming structures do not include the $AF$ state.}
 		\label{fig:fcg}
 	\end{figure}

 Tunability of $p$ and $q$ allows us to observe many more spin domain formation, where the $AF$ phase does not participate. At $q=30 Hz$, if $p$ is relaxed to a moderate negative value, a domain formation between the $MF2$ state, residing at the centre and $F1$ staying outside can be observed (Fig~\ref{fig:fcg}\subref{subfig-1:q30cg3}). A simple check, as mentioned earlier, can be helpful to see that $MF2$ can indeed appear in this parameter domain. Similarly, at $p=100 Hz$, tunability of $q$ around $48 Hz$ would let one observe the condensate forming a domain structure with $MF1$ inside and $F2$ outside. The magnitude of $q$ is roughly in the same region as compared to (Fig~\ref{fig:fcg}\subref{subfig-1:q30cg3}) but $p$ is positive in this case (not shown in Fig~\ref{fig:fcg}).
 When $p$ is fixed at $100 Hz$, variation around small negative values of quadratic term reveals that a domain structure between the two ferromagnetic phase can be observed (Fig~\ref{fig:fcg}\subref{subfig-2:p100cg2}). Note that, this type of structure is not possible for untrapped situation (Fig~\ref{fig:hom}\subref{subfig-1:hom1}), which reveals the novelty of the trapped condensate. For relatively smaller $p$, variation of $q$ around $30.5 Hz$ favors the $PM$ state energetically to occupy the high density region. $F2$ becomes the most stable state to capture the low density region (Fig~\ref{fig:fcg}\subref{subfig-3:p30cg1}).
 Obviously as $|q|>|p|$, $(1,1,1)$ can be identified as a $PM$ state. The same structure extends to larger values of $p$ ($=100 Hz$) and $q$ ($101.5$ to $104 Hz$) which is not shown in the figure. Fig~\ref{fig:fcg}\subref{subfig-4:q100cg3} draws one's attention to compare it with  Fig~\ref{fig:fcg}\subref{subfig-3:p30cg1}. Though the parameter domain in this case is different but similar structure with $PM$ state inside and a ferromagnetic state (in this case $F1$) outside is observed.\\
 \par
  Comparison between Fig~\ref{fig:fcg}\subref{subfig-5:p100cg1} and \subref{subfig-4:q100cg3} may be enough to draw a deceptive conclusion that the change in signs of $p$ and $q$ leads to the change of position between the states $F1$ and $PM$ inside the trap. One should notice that the $(1,1,1)$ state actually corresponds to $APM$ state as $|p|>|q|$ here. When $q$ is at around the same region (say at $-100Hz$) and $p$ is also in the same region in the negative half (around $-100.5 Hz$), a same type of structure is observed with $F2$ in place of $F1$ staying at the trap core (not shown in the figure). In the parametric domain described in Fig~\ref{fig:fcg}\subref{subfig-6:qm5cg1} the mixed ferromagnetic domain $MF1$ gains the higher density region of the trap while the $APM$ state gains the lower density region.

	\begin{figure}[h!]
		\subfloat[$c_1>0 $\label{subfig-1:p5cg1}]{%
			\includegraphics[width=0.24\textwidth]{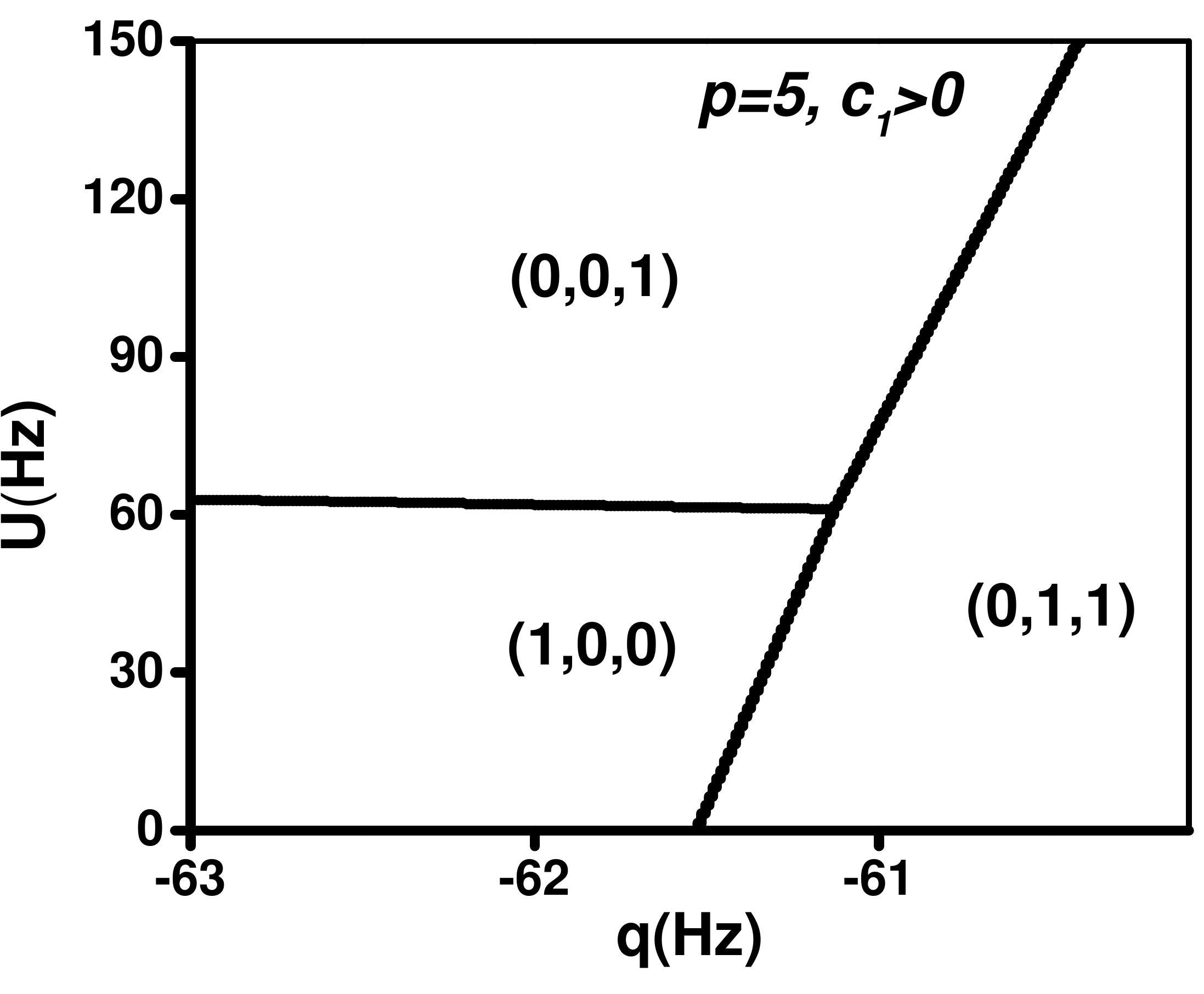}
		}
		\subfloat[$c_1>0 $\label{subfig-2:qm5cg3}]{%
			\includegraphics[width=0.24\textwidth]{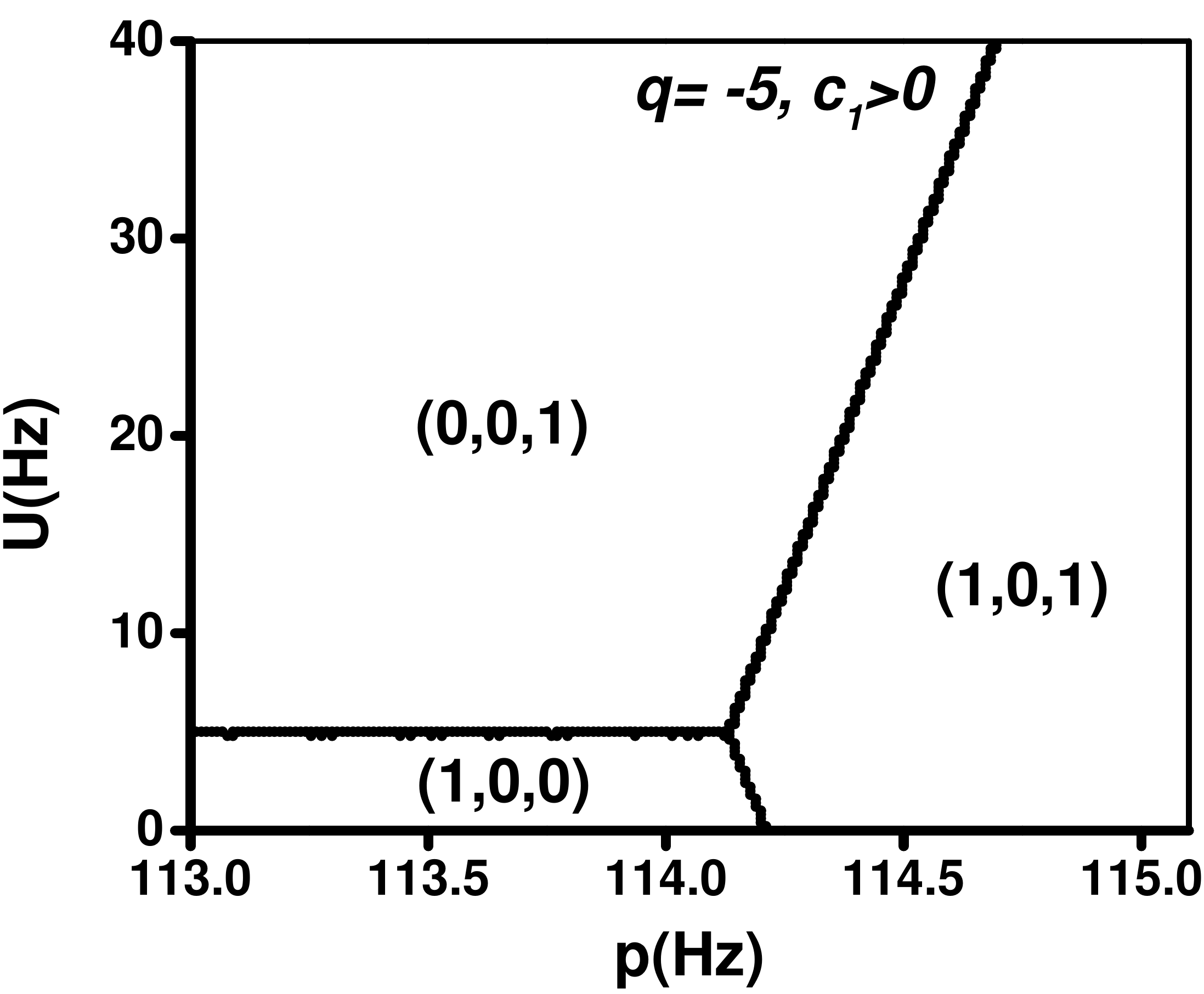}
		}
			
		\subfloat[$c_1>0 $\label{subfig-3:qm5cg2}]{%
			\includegraphics[width=0.24\textwidth]{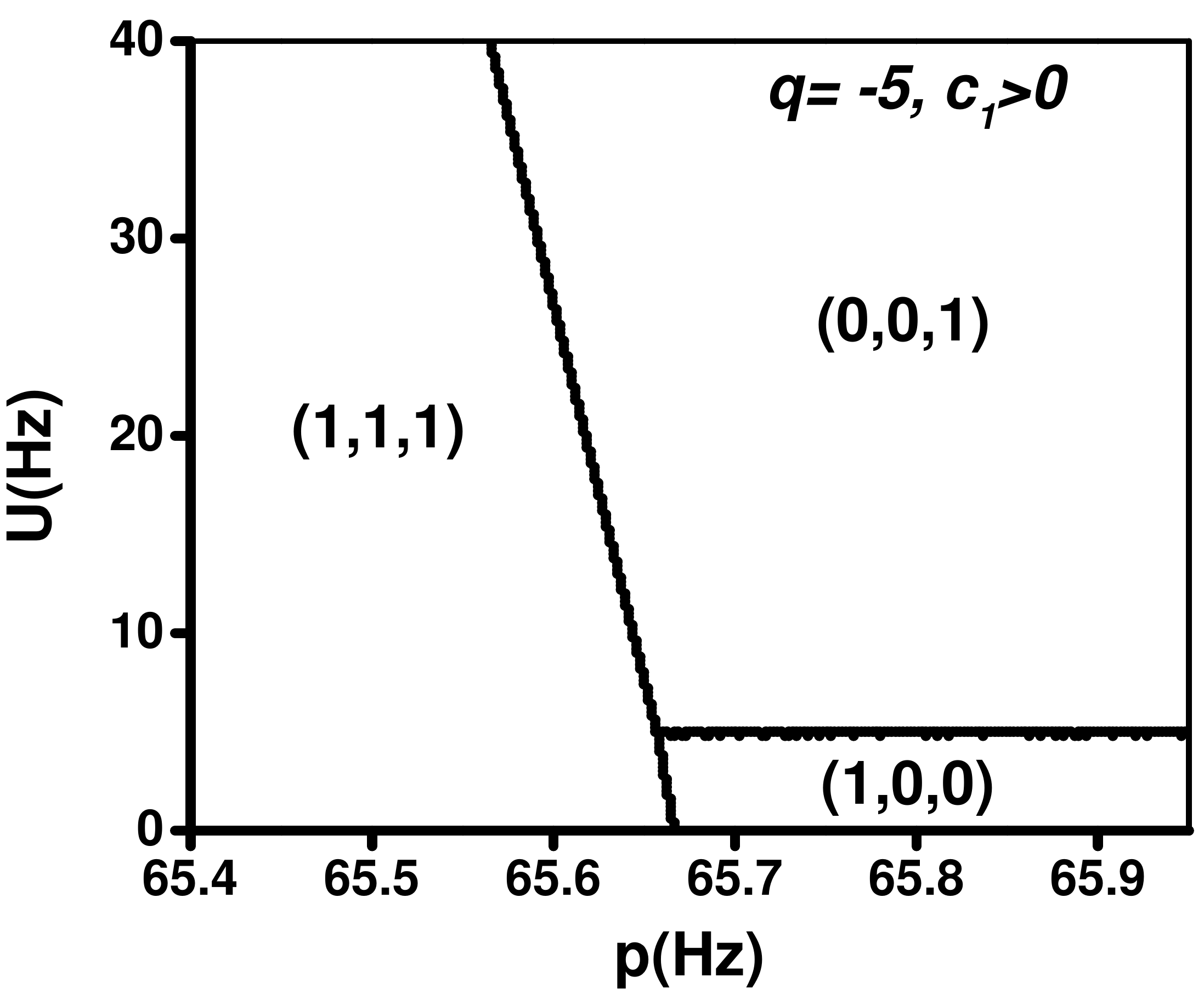}
		}
	    \subfloat[$c_1>0 $\label{subfig-4:qm5cg4}]{%
			\includegraphics[width=0.24\textwidth]{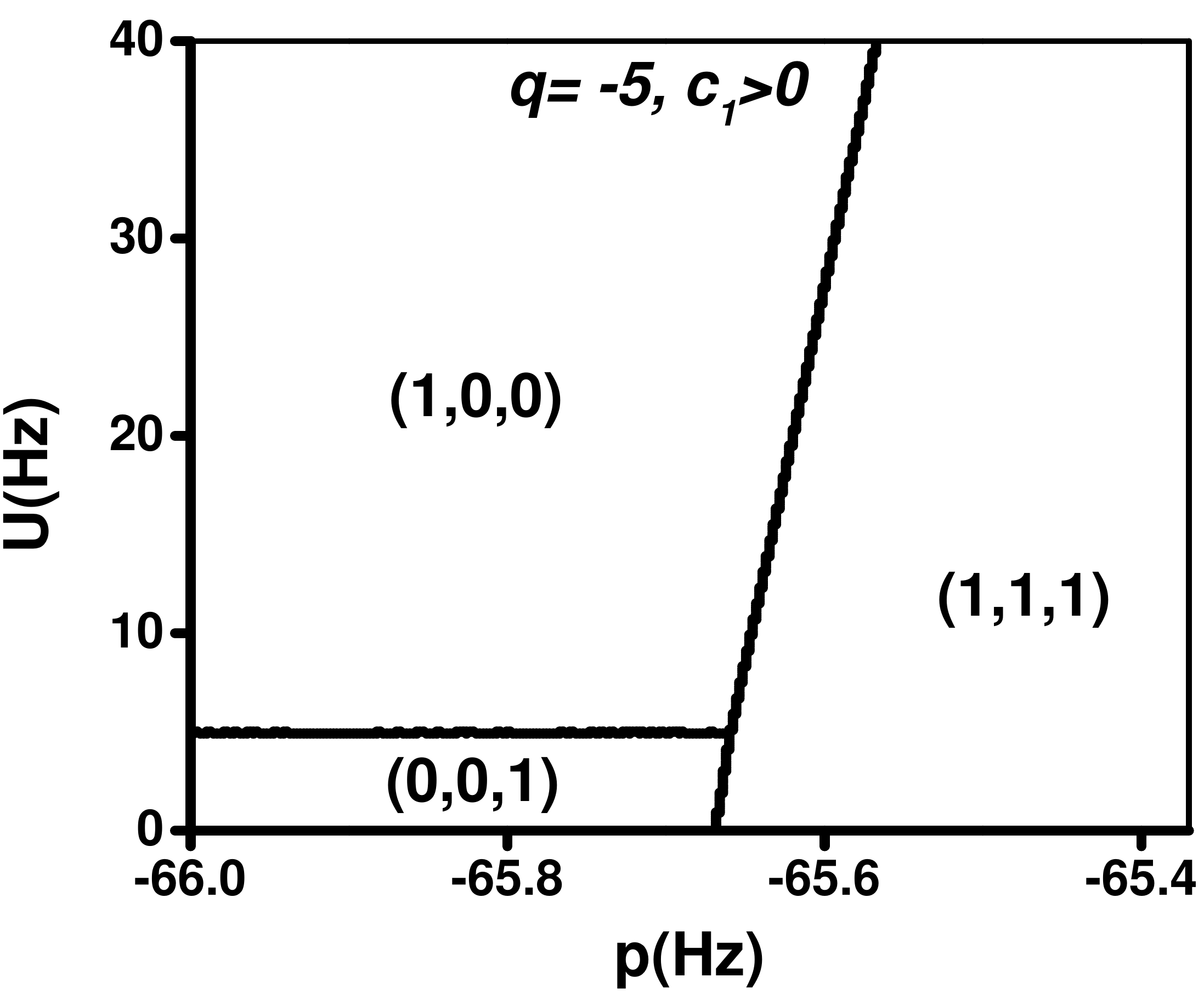}
		}
	
   		\centering
   		\subfloat[$c_1>0 $\label{subfig-5:qm100cg2}]{%
        	\includegraphics[width=0.24\textwidth]{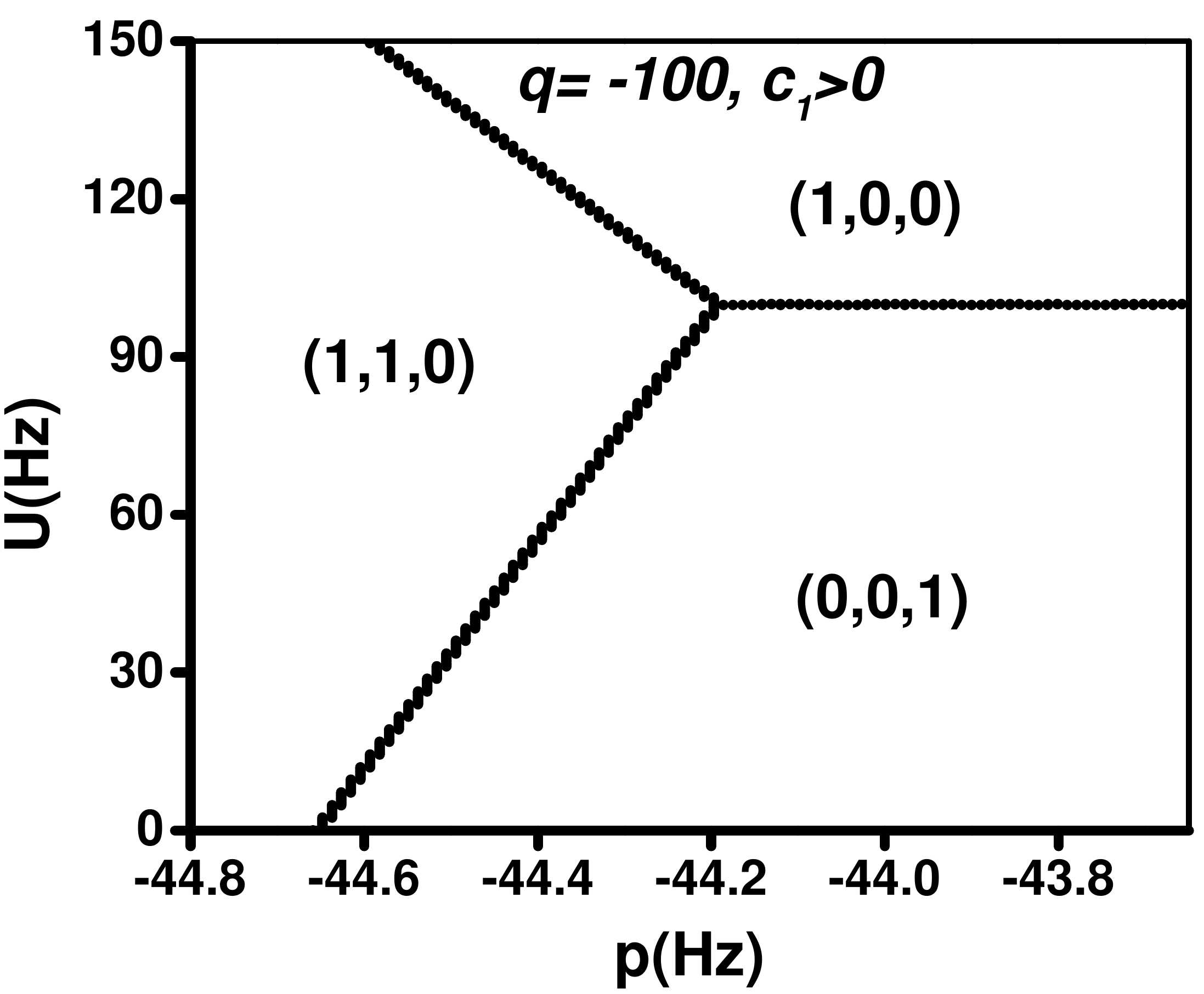}
        }	
		\caption{Domain formation possibilities with three coexisting states for different values of $p$ and $q$. Note that all these possibilities arise when $q$ assumes a negative value.}
		\label{fig:tripple}
	\end{figure}

\par
Fig~\ref{fig:tripple} summarizes all the various possibilities of coexistence of three states that we have observed. Setting the linear term $p$ to a small positive value, say $5 Hz$, opens up the possibility to observe a domain formation of three states when $q$ is tuned at around $-61.5 Hz$ (Fig~\ref{fig:tripple}\subref{subfig-1:p5cg1}). The mixed ferromagnetic phase $MF2$ outplays all other states to stay at the low potential region. At a distance from the trap centre $F1$ appears as it becomes the lowest energy state. Drawing an imaginary vertical line one can find the corresponding $U(\vec{r})$, where the first PS happens, in turn allowing to find the domain of $MF2$. For 2D harmonic trap the previously defined $r_0$ becomes, $r_0=\sqrt{2U/\omega}$. Following the same scheme it is easy to find out the distance from the centre at which the next state $F2$ resides. Note that $(p+q)$ being negative here, does not impose any restriction over the existence of the state $MF2$ (see Table II).
\par
For large values of $p$ around $p=114.20 Hz$ and small negative $q$(=$-5Hz$), another three layer domain formation can be observed(Fig~\ref{fig:tripple}\subref{subfig-2:qm5cg3}). Here the state $F2$ is only allowed to form in the most exterior part of the trap. The other ferromagnetic state $(1,0,0)$ gets the central region and the $AF$ state separates them. Tuning $p$ to a lower value while keeping $q$ fixed, a different structure can be seen (Fig~\ref{fig:tripple}\subref{subfig-3:qm5cg2}) when at an approximate $p=65.66 Hz$ the $F2$ state still remains at a furthest distance from the centre, but APM phase occupies the higher density region. $(1,0,0)$ wins energetically to occupy the stay in between them.
\par
Comparison between Fig~\ref{fig:tripple}\subref{subfig-3:qm5cg2} and Fig~\ref{fig:tripple}\subref{subfig-4:qm5cg4} dictates the role of $p$ on the appearance of the ferromagnetic state. For moderately small negative $p$, when $q$ is largely negative, a domain structure of two ferromagnetic state and the $MF1$ can be observed with $F2$ at the core and $F1$ in the most outer region are separated by a layer of $MF1$.
\subsubsection{\underline{3.2.2 Ferromagnetic interaction $c_1< 0$}}\label{subsubsec:cless}
\begin{figure}[h!]
		\subfloat[$c_1<0 $\label{subfig-1:p50cn1}]{%
			\includegraphics[width=0.24\textwidth]{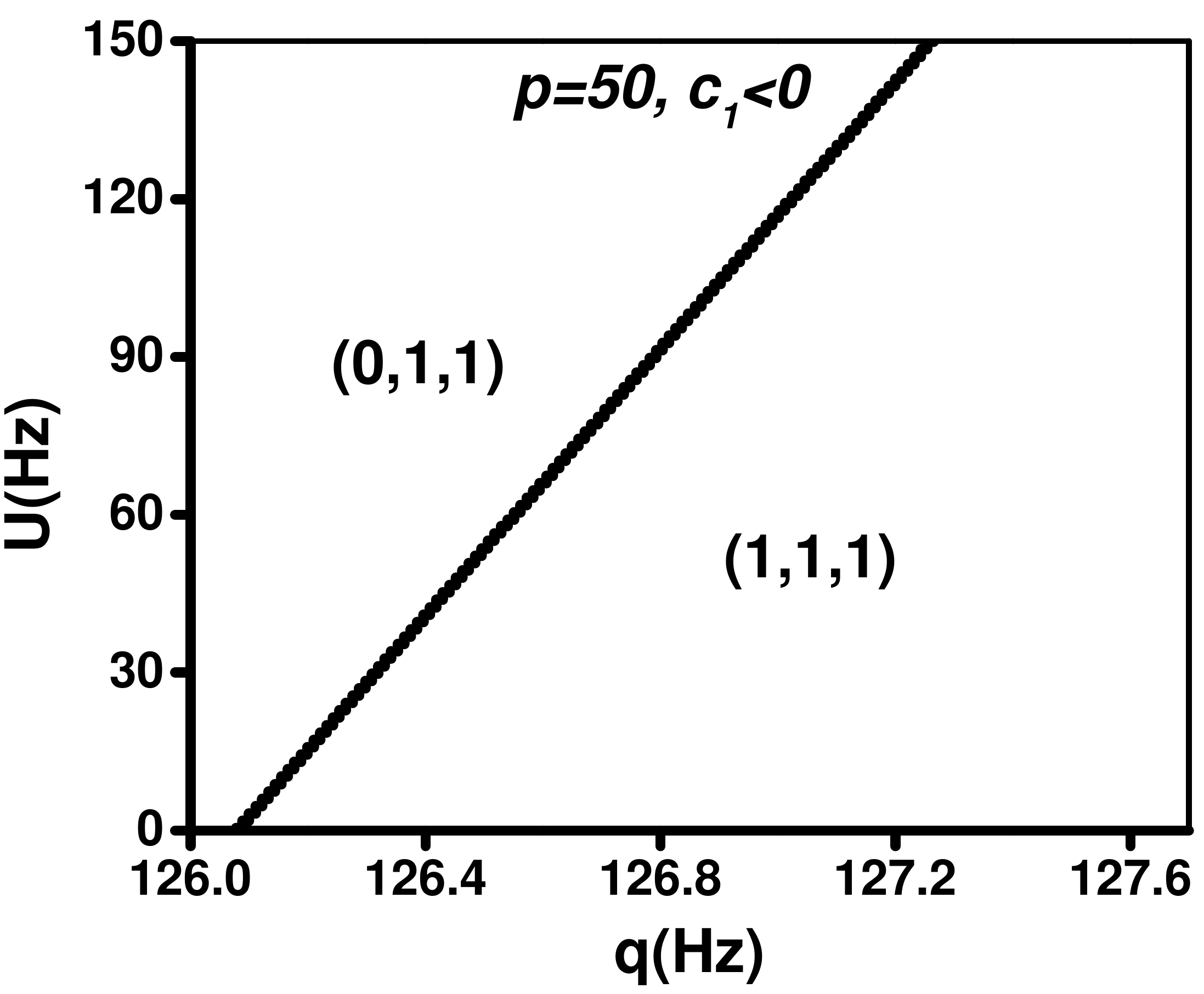}
		}
		\subfloat[$c_1<0 $\label{subfig-2:p5cn1}]{%
			\includegraphics[width=0.24\textwidth]{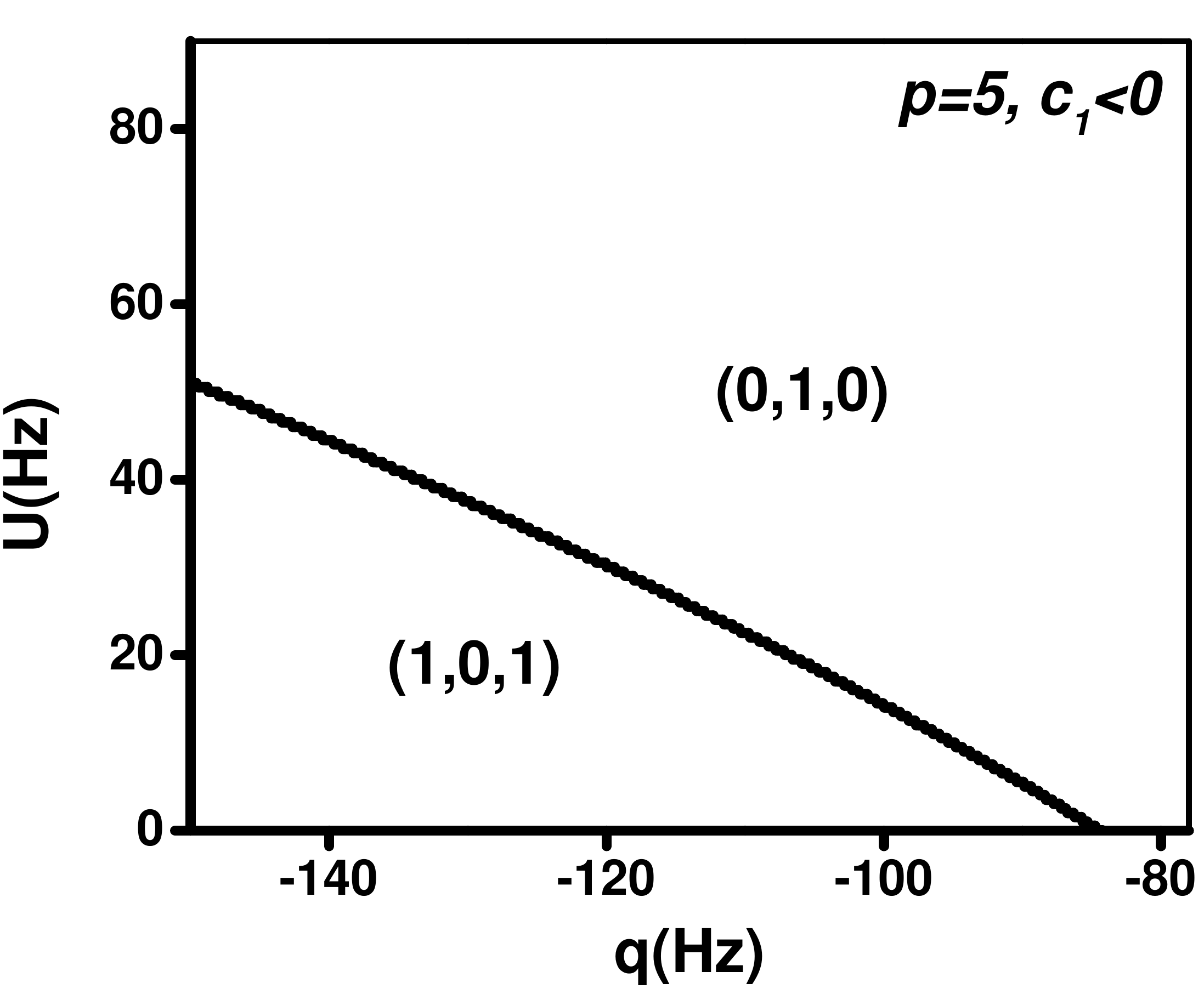}
		}
	
		\subfloat[$c_1<0 $\label{subfig-3:q5cn2}]{%
			\includegraphics[width=0.24\textwidth]{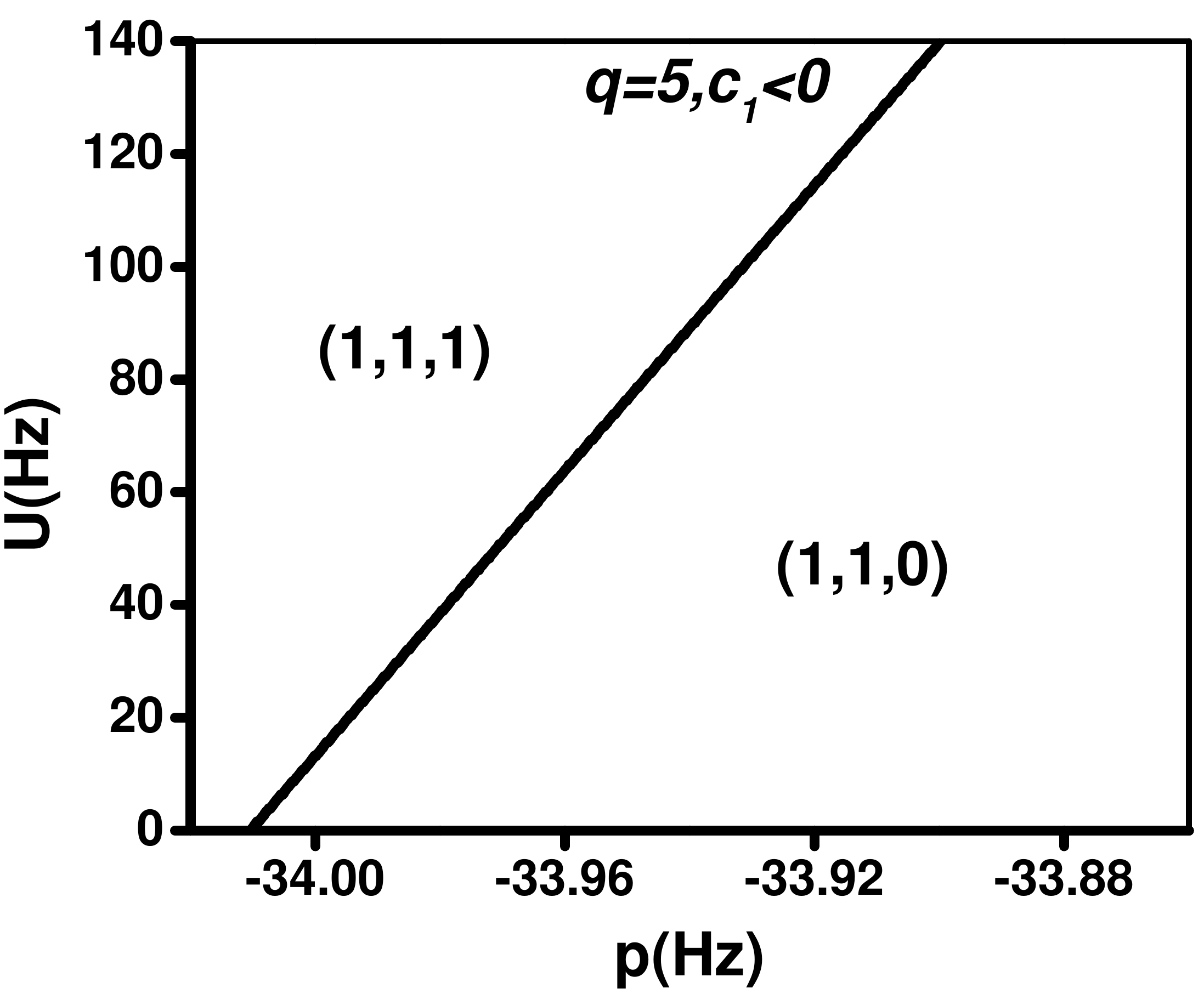}
		}
	    \subfloat[$c_1<0 $\label{subfig-4:q5cn1}]{%
			\includegraphics[width=0.24\textwidth]{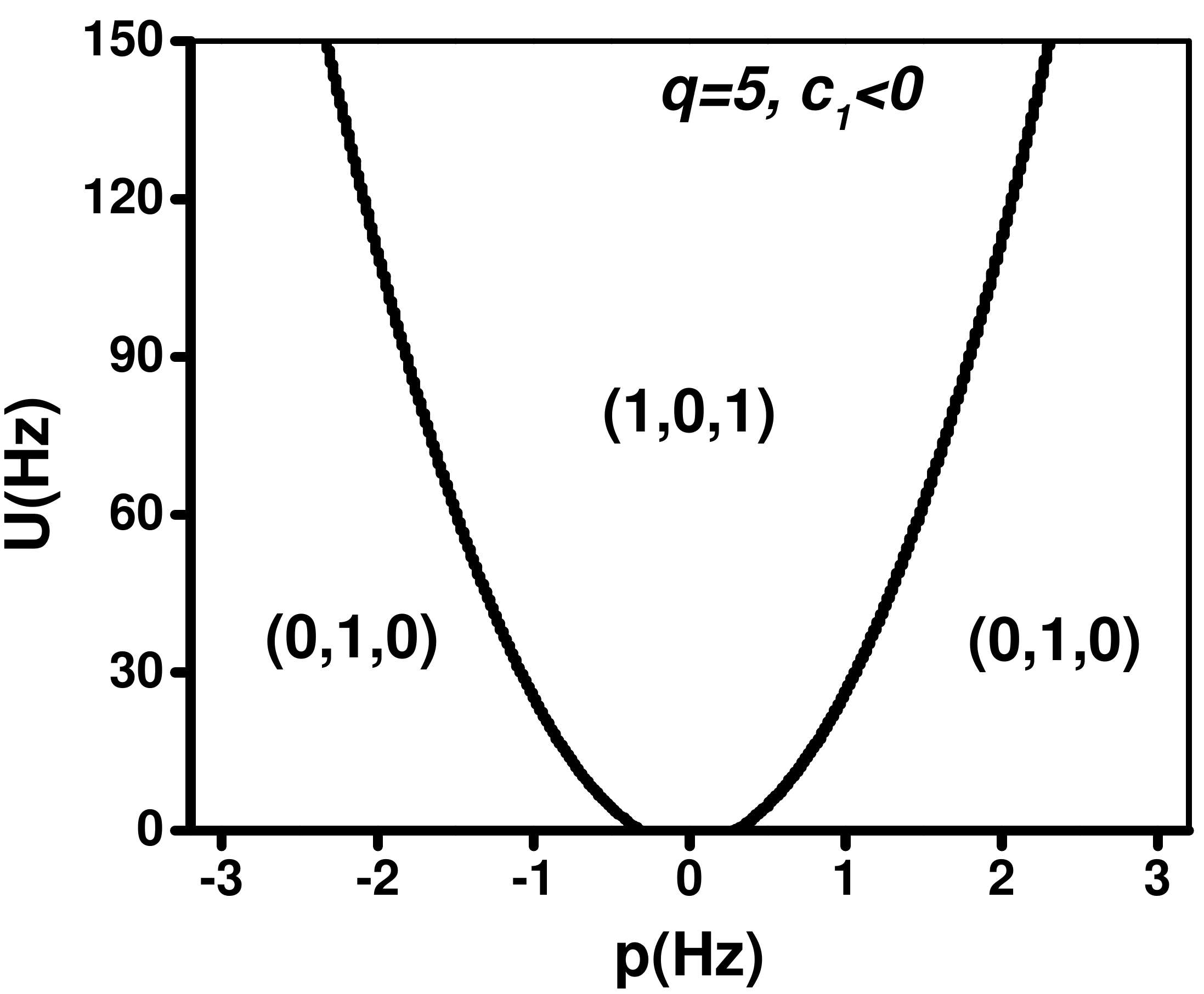}
		}
	
		\centering	
		\subfloat[$c_1<0 $\label{subfig-5:qm5cn1}]{%
			\includegraphics[width=0.24\textwidth]{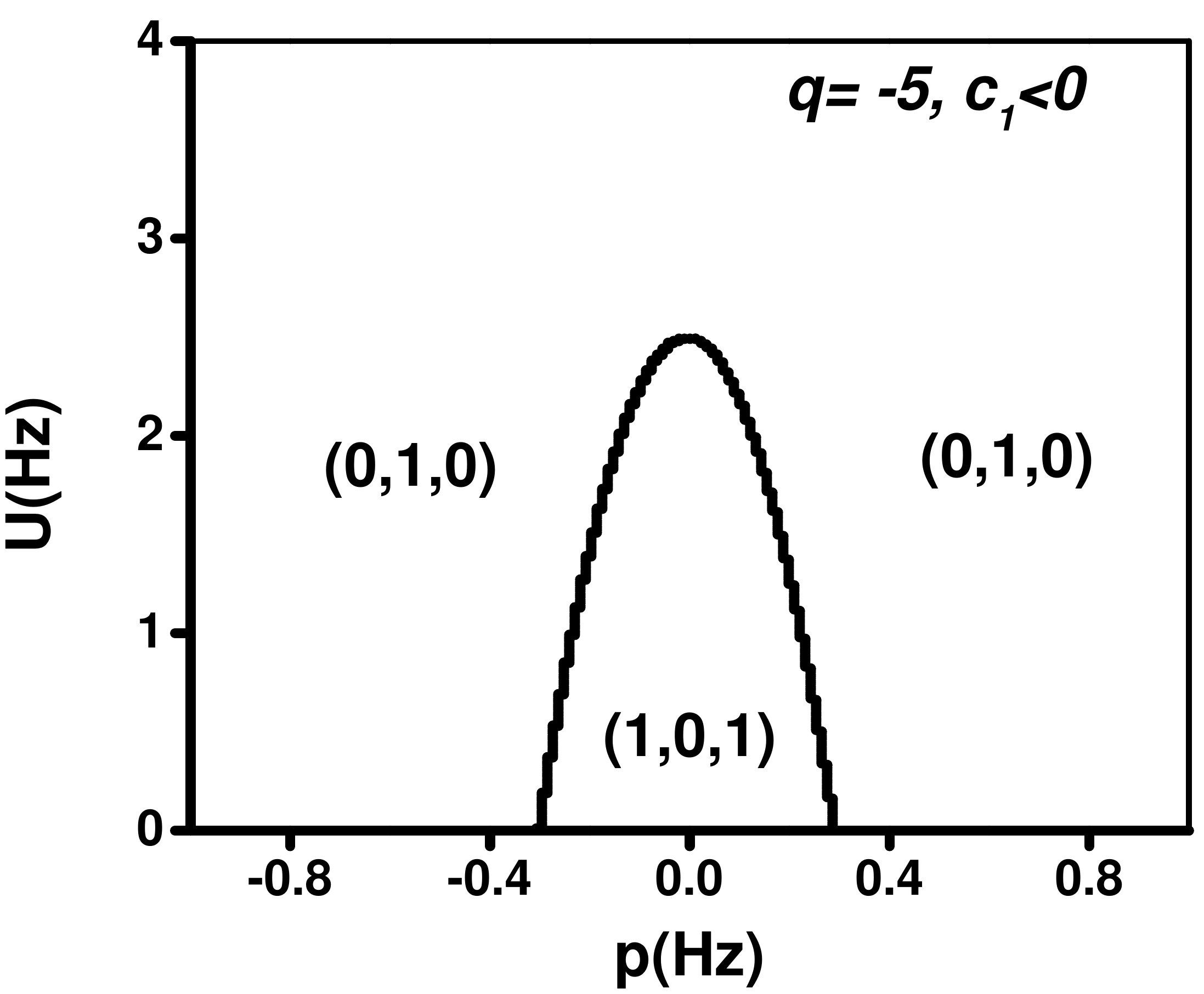}
		}
		
		\caption{All possible phase separation for ferromagnetic type interaction.}
		\label{fig:allcn}
	\end{figure}

 To investigate the domain formation phenomenon for ferromagnetic type of interaction we choose $^{87}Rb$ for which $c_1$ comes out to be $-0.275\times 10^{-19} Hz$. The parameter $c_0$ is numerically $78.02 \times 10^{-19} Hz$ for this element. Again all the controllable parameter and the trapping potential is varied over the specified range as stated in subsection 3.2.1. The $\mu$ is kept fixed at the value mentioned earlier. We find that a relatively small amount of domain formation can happen for the ferromagnetic type of interaction as compared to the plethora of structures seen in the last subsection. The most startling fact here is that there is no dominance of the ferromagnetic phases in the domain formation scenario as observed in this parameter region. Note that there is no apparent reason for the ferromagnetic state not to appear at all parameter regime; in fact we found out that in an extended parameter region (tuning $p$ and $q$ beyond $\pm150 Hz$) ferromagnetic state dominates in the domain formation scenario. As we are restricting ourselves in the parameter region discussed above, we are not including these cases in Fig~\ref{fig:allcn}.

 \par
We first fix the value of the linear Zeeman term. In the parameter region as shown in Fig~\ref{fig:allcn}\subref{subfig-1:p50cn1}, the $MF2$ state is the most stable one to prevail at the outer region of the trap while the PM state stays at the core. A slight increment in the $q$ value would only result in the broadening of the PM domain. For small $p$ (Fig~\ref{fig:allcn}\subref{subfig-2:p5cn1}), the state $(0,1,0)$ often called the polar state appears to have a phase separating structure with the $AF$ state. Note that, this happens at a large negative value of $q$. By fixing the quadratic term at a small positive value like $5 Hz$ and tuning the linear Zeeman term to a relatively moderate negative value allows one to see another structure between $MF1$ and the $APM$ state (Fig~\ref{fig:allcn}\subref{subfig-3:q5cn2}). For a nonzero small value of $|p|$ the polar state stays central followed by the $AF$ state staying wide(Fig~\ref{fig:allcn}\subref{subfig-4:q5cn1}).

As the sign of $q$ is changed, a comparison between Fig~\ref{fig:allcn}\subref{subfig-5:qm5cn1} and Fig~\ref{fig:allcn}\subref{subfig-4:q5cn1} reveals the interchange of the domains of polar and $AF$ state. In this case the $AF$ state forms at the centre. A slight increment of $|p|$ would prefer the polar phase to expand its domain in both the cases. In this case after a limiting value of $|p|$ the structure is lost. Interesting to note that as $p$ appears in the energy expression of the $AF$ state (for details see the Table-II) an increment in $p^2$ would increase the energy density of it for $c_1<0$. As $p$ does not appear in the energy density expression of the polar state, depletion of the domain of the $AF$ is quite reasonable to occur.
\par
Note that, Eq~\ref{eq:rev7} suggests the quadratic term does not appear in the expression of energy density of the polar phase (as $m=0$ is only present) but the $AF$ state gets affected approximately as $qn$ ($n$ being the total number density). Therefore, a change in sign of $q$ from positive (Fig~\ref{fig:allcn}\subref{subfig-4:q5cn1}) to negative (Fig~\ref{fig:allcn}\subref{subfig-5:qm5cn1}) can only decrease the energy density of the $AF$ state, thus allowing it to be energetically more stable at the high density region.
 
\section{4.Discussion}
Using T-F approximation, we have studied the phase separation of stationary states in details for a spin-1 condensate with both ferromagnetic and anti-ferromagnetic type of interaction. We show here that this procedure is indeed very general and can capture all the mixed phases equally, irrespective of the confining potential. However, the test case that has been considered in the present work makes use of an isotropic harmonic confinement. Applying optical and magnetic Feshbach resonance \cite{PhysRevA.51.4852}, the spin interaction parameter can be tuned \cite{inouye1998observation,RevModPhys.82.1225} close to zero \cite{PhysRevA.92.023616}. For this case also, all the possible potential induced domain structure has been investigated here in details. 
\par
It should be noted that the Zeeman terms may be varied to even higher values \cite{PhysRevLett.119.050404} and the scheme shown here using energy density comparison should suffice to reveal any domain structure even in that regime. At zero magnetic field the system becomes degenerate even at non-zero temperature \cite{PhysRevLett.119.050404}. Careful observation reveals that the energy density corresponding to the $(A)PM$ state is ill defined at $q=0$ (in Table-II). To get rid of this issue, one can rewrite Eq~\ref{eq:rev12},\ref{eq:rev13} for $p,q=0$ in the first place. It is easy to see that the subcomponent density then would be multiples of each other, a fact which agrees with the assumption taken by S. Gautam and S.K. Adhikari the article \cite{PhysRevA.92.023616}.
\par 
The broader picture of phase separation in terms of stationary states presented in this paper is the first step and many of the situations arising may get ruled out when a stability analysis is done with respect to density and the phase perturbations. However, this picture is essential in order to know in the beginning about all the equilibrium possibilities, that can exist and following this, particular stability analysis should happen. The present analysis is quite interesting in that respect because it shows that a complete treatment of all the phases on equal footing is possible under T-F approximation, which to our knowledge has not been done in this way in existing literature, but, has been done in bits and pieces in many papers. 
\par
The present analysis also shows that the actual phase boundaries over space can be obtained in the T-F approximation and the next natural step could be looking at the dynamics of those under various conditions and perturbations. The constant chemical potential constraint which is essential for chemical stability of coexisting phases may also be a heavy requirement for many cases under various conditions and the failure of maintaining this constraint may also rule out some otherwise allowed structures. However, that can only be understood once we look at into the requirement of fulfillment of this constraint on the other basis and thereby restrict this broader picture systematically. Nevertheless, even for all these, the broader picture of stationary phase separated domains is necessary and the present analysis will help in that purpose.
\section{Acknowledgment}
 PKK would like to thank the Council of Scientific and Industrial Research (CSIR), India for the funding provided and Sourav Laha for many helpful discussion.
\bibliographystyle{apsrev4-1}
\bibliography{reference1}

\def\germ{\frak} \def\scr{\cal} \ifx\documentclass\undefinedcs
  \def\bf{\fam\bffam\tenbf}\def\rm{\fam0\tenrm}\fi 
  \def\defaultdefine#1#2{\expandafter\ifx\csname#1\endcsname\relax
  \expandafter\def\csname#1\endcsname{#2}\fi} \defaultdefine{Bbb}{\bf}
  \defaultdefine{frak}{\bf} \defaultdefine{=}{\B} 
  \defaultdefine{mathfrak}{\frak} \defaultdefine{mathbb}{\bf}
  \defaultdefine{mathcal}{\cal}
  \defaultdefine{beth}{BETH}\defaultdefine{cal}{\bf} \def\bbfI{{\Bbb I}}
  \def\mbox{\hbox} \def\text{\hbox} \def\om{\omega} \def\Cal#1{{\bf #1}}
  \def\pcf{pcf} \defaultdefine{cf}{cf} \defaultdefine{reals}{{\Bbb R}}
  \defaultdefine{real}{{\Bbb R}} \def\restriction{{|}} \def\club{CLUB}
  \def\w{\omega} \def\exist{\exists} \def\se{{\germ se}} \def\bb{{\bf b}}
  \def\equivalence{\equiv} \let\lt< \let\gt>
\begin{thebibliography}{33}%
\makeatletter
\providecommand \@ifxundefined [1]{%
 \@ifx{#1\undefined}
}%
\providecommand \@ifnum [1]{%
 \ifnum #1\expandafter \@firstoftwo
 \else \expandafter \@secondoftwo
 \fi
}%
\providecommand \@ifx [1]{%
 \ifx #1\expandafter \@firstoftwo
 \else \expandafter \@secondoftwo
 \fi
}%
\providecommand \natexlab [1]{#1}%
\providecommand \enquote  [1]{``#1''}%
\providecommand \bibnamefont  [1]{#1}%
\providecommand \bibfnamefont [1]{#1}%
\providecommand \citenamefont [1]{#1}%
\providecommand \href@noop [0]{\@secondoftwo}%
\providecommand \href [0]{\begingroup \@sanitize@url \@href}%
\providecommand \@href[1]{\@@startlink{#1}\@@href}%
\providecommand \@@href[1]{\endgroup#1\@@endlink}%
\providecommand \@sanitize@url [0]{\catcode `\\12\catcode `\$12\catcode
  `\&12\catcode `\#12\catcode `\^12\catcode `\_12\catcode `\%12\relax}%
\providecommand \@@startlink[1]{}%
\providecommand \@@endlink[0]{}%
\providecommand \url  [0]{\begingroup\@sanitize@url \@url }%
\providecommand \@url [1]{\endgroup\@href {#1}{\urlprefix }}%
\providecommand \urlprefix  [0]{URL }%
\providecommand \Eprint [0]{\href }%
\providecommand \doibase [0]{http://dx.doi.org/}%
\providecommand \selectlanguage [0]{\@gobble}%
\providecommand \bibinfo  [0]{\@secondoftwo}%
\providecommand \bibfield  [0]{\@secondoftwo}%
\providecommand \translation [1]{[#1]}%
\providecommand \BibitemOpen [0]{}%
\providecommand \bibitemStop [0]{}%
\providecommand \bibitemNoStop [0]{.\EOS\space}%
\providecommand \EOS [0]{\spacefactor3000\relax}%
\providecommand \BibitemShut  [1]{\csname bibitem#1\endcsname}%
\let\auto@bib@innerbib\@empty
\bibitem [{\citenamefont {Timmermans}(1998)}]{PhysRevLett.81.5718}%
  \BibitemOpen
  \bibfield  {author} {\bibinfo {author} {\bibfnamefont {E.}~\bibnamefont
  {Timmermans}},\ }\href {\doibase 10.1103/PhysRevLett.81.5718} {\bibfield
  {journal} {\bibinfo  {journal} {Phys. Rev. Lett.}\ }\textbf {\bibinfo
  {volume} {81}},\ \bibinfo {pages} {5718} (\bibinfo {year}
  {1998})}\BibitemShut {NoStop}%
\bibitem [{\citenamefont {Stenger}\ \emph {et~al.}(1998)\citenamefont
  {Stenger}, \citenamefont {Inouye}, \citenamefont {Stamper-Kurn},
  \citenamefont {Miesner}, \citenamefont {Chikkatur},\ and\ \citenamefont
  {Ketterle}}]{stenger98}%
  \BibitemOpen
  \bibfield  {author} {\bibinfo {author} {\bibfnamefont {J.}~\bibnamefont
  {Stenger}}, \bibinfo {author} {\bibfnamefont {S.}~\bibnamefont {Inouye}},
  \bibinfo {author} {\bibfnamefont {D.}~\bibnamefont {Stamper-Kurn}}, \bibinfo
  {author} {\bibfnamefont {H.-J.}\ \bibnamefont {Miesner}}, \bibinfo {author}
  {\bibfnamefont {A.}~\bibnamefont {Chikkatur}}, \ and\ \bibinfo {author}
  {\bibfnamefont {W.}~\bibnamefont {Ketterle}},\ }\href@noop {} {\bibfield
  {journal} {\bibinfo  {journal} {Nature}\ }\textbf {\bibinfo {volume} {396}},\
  \bibinfo {pages} {345} (\bibinfo {year} {1998})}\BibitemShut {NoStop}%
\bibitem [{\citenamefont {Isoshima}\ \emph {et~al.}(1999)\citenamefont
  {Isoshima}, \citenamefont {Machida},\ and\ \citenamefont
  {Ohmi}}]{PhysRevA.60.4857}%
  \BibitemOpen
  \bibfield  {author} {\bibinfo {author} {\bibfnamefont {T.}~\bibnamefont
  {Isoshima}}, \bibinfo {author} {\bibfnamefont {K.}~\bibnamefont {Machida}}, \
  and\ \bibinfo {author} {\bibfnamefont {T.}~\bibnamefont {Ohmi}},\ }\href
  {\doibase 10.1103/PhysRevA.60.4857} {\bibfield  {journal} {\bibinfo
  {journal} {Phys. Rev. A}\ }\textbf {\bibinfo {volume} {60}},\ \bibinfo
  {pages} {4857} (\bibinfo {year} {1999})}\BibitemShut {NoStop}%
\bibitem [{\citenamefont {Matuszewski}(2010)}]{mats10gr}%
  \BibitemOpen
  \bibfield  {author} {\bibinfo {author} {\bibfnamefont {M.}~\bibnamefont
  {Matuszewski}},\ }\href {\doibase 10.1103/PhysRevA.82.053630} {\bibfield
  {journal} {\bibinfo  {journal} {Phys. Rev. A}\ }\textbf {\bibinfo {volume}
  {82}},\ \bibinfo {pages} {053630} (\bibinfo {year} {2010})}\BibitemShut
  {NoStop}%
\bibitem [{\citenamefont {Matuszewski}\ \emph {et~al.}(2008)\citenamefont
  {Matuszewski}, \citenamefont {Alexander},\ and\ \citenamefont
  {Kivshar}}]{PhysRevA.78.023632}%
  \BibitemOpen
  \bibfield  {author} {\bibinfo {author} {\bibfnamefont {M.}~\bibnamefont
  {Matuszewski}}, \bibinfo {author} {\bibfnamefont {T.~J.}\ \bibnamefont
  {Alexander}}, \ and\ \bibinfo {author} {\bibfnamefont {Y.~S.}\ \bibnamefont
  {Kivshar}},\ }\href {\doibase 10.1103/PhysRevA.78.023632} {\bibfield
  {journal} {\bibinfo  {journal} {Phys. Rev. A}\ }\textbf {\bibinfo {volume}
  {78}},\ \bibinfo {pages} {023632} (\bibinfo {year} {2008})}\BibitemShut
  {NoStop}%
\bibitem [{\citenamefont {Matuszewski}\ \emph {et~al.}(2009)\citenamefont
  {Matuszewski}, \citenamefont {Alexander},\ and\ \citenamefont
  {Kivshar}}]{PhysRevA.80.023602}%
  \BibitemOpen
  \bibfield  {author} {\bibinfo {author} {\bibfnamefont {M.}~\bibnamefont
  {Matuszewski}}, \bibinfo {author} {\bibfnamefont {T.~J.}\ \bibnamefont
  {Alexander}}, \ and\ \bibinfo {author} {\bibfnamefont {Y.~S.}\ \bibnamefont
  {Kivshar}},\ }\href {\doibase 10.1103/PhysRevA.80.023602} {\bibfield
  {journal} {\bibinfo  {journal} {Phys. Rev. A}\ }\textbf {\bibinfo {volume}
  {80}},\ \bibinfo {pages} {023602} (\bibinfo {year} {2009})}\BibitemShut
  {NoStop}%
\bibitem [{\citenamefont {\ifmmode~\acute{S}\else \'{S}\fi{}wis\l{}ocki}\ and\
  \citenamefont {Matuszewski}(2012)}]{PhysRevA.85.023601}%
  \BibitemOpen
  \bibfield  {author} {\bibinfo {author} {\bibfnamefont {T.}~\bibnamefont
  {\ifmmode~\acute{S}\else \'{S}\fi{}wis\l{}ocki}}\ and\ \bibinfo {author}
  {\bibfnamefont {M.}~\bibnamefont {Matuszewski}},\ }\href {\doibase
  10.1103/PhysRevA.85.023601} {\bibfield  {journal} {\bibinfo  {journal} {Phys.
  Rev. A}\ }\textbf {\bibinfo {volume} {85}},\ \bibinfo {pages} {023601}
  (\bibinfo {year} {2012})}\BibitemShut {NoStop}%
\bibitem [{\citenamefont {Gautam}\ and\ \citenamefont
  {Adhikari}(2015)}]{PhysRevA.92.023616}%
  \BibitemOpen
  \bibfield  {author} {\bibinfo {author} {\bibfnamefont {S.}~\bibnamefont
  {Gautam}}\ and\ \bibinfo {author} {\bibfnamefont {S.~K.}\ \bibnamefont
  {Adhikari}},\ }\href {\doibase 10.1103/PhysRevA.92.023616} {\bibfield
  {journal} {\bibinfo  {journal} {Phys. Rev. A}\ }\textbf {\bibinfo {volume}
  {92}},\ \bibinfo {pages} {023616} (\bibinfo {year} {2015})}\BibitemShut
  {NoStop}%
\bibitem [{\citenamefont {Jim{\'e}nez-Garc{\'\i}a}\ \emph
  {et~al.}(2018)\citenamefont {Jim{\'e}nez-Garc{\'\i}a}, \citenamefont
  {Invernizzi}, \citenamefont {Evrard}, \citenamefont {Frapolli}, \citenamefont
  {Dalibard},\ and\ \citenamefont {Gerbier}}]{jimenez2018spontaneous}%
  \BibitemOpen
  \bibfield  {author} {\bibinfo {author} {\bibfnamefont {K.}~\bibnamefont
  {Jim{\'e}nez-Garc{\'\i}a}}, \bibinfo {author} {\bibfnamefont
  {A.}~\bibnamefont {Invernizzi}}, \bibinfo {author} {\bibfnamefont
  {B.}~\bibnamefont {Evrard}}, \bibinfo {author} {\bibfnamefont
  {C.}~\bibnamefont {Frapolli}}, \bibinfo {author} {\bibfnamefont
  {J.}~\bibnamefont {Dalibard}}, \ and\ \bibinfo {author} {\bibfnamefont
  {F.}~\bibnamefont {Gerbier}},\ }\href@noop {} {\bibfield  {journal} {\bibinfo
   {journal} {arXiv preprint arXiv:1808.01015}\ } (\bibinfo {year}
  {2018})}\BibitemShut {NoStop}%
\bibitem [{\citenamefont {Ho}\ and\ \citenamefont
  {Shenoy}(1996)}]{PhysRevLett.77.3276}%
  \BibitemOpen
  \bibfield  {author} {\bibinfo {author} {\bibfnamefont {T.-L.}\ \bibnamefont
  {Ho}}\ and\ \bibinfo {author} {\bibfnamefont {V.~B.}\ \bibnamefont
  {Shenoy}},\ }\href {\doibase 10.1103/PhysRevLett.77.3276} {\bibfield
  {journal} {\bibinfo  {journal} {Phys. Rev. Lett.}\ }\textbf {\bibinfo
  {volume} {77}},\ \bibinfo {pages} {3276} (\bibinfo {year}
  {1996})}\BibitemShut {NoStop}%
\bibitem [{\citenamefont {Sabbatini}\ \emph {et~al.}(2011)\citenamefont
  {Sabbatini}, \citenamefont {Zurek},\ and\ \citenamefont
  {Davis}}]{sabbatini2011phase}%
  \BibitemOpen
  \bibfield  {author} {\bibinfo {author} {\bibfnamefont {J.}~\bibnamefont
  {Sabbatini}}, \bibinfo {author} {\bibfnamefont {W.~H.}\ \bibnamefont
  {Zurek}}, \ and\ \bibinfo {author} {\bibfnamefont {M.~J.}\ \bibnamefont
  {Davis}},\ }\href {\doibase 10.1103/PhysRevLett.107.230402} {\bibfield
  {journal} {\bibinfo  {journal} {Phys. Rev. Lett.}\ }\textbf {\bibinfo
  {volume} {107}},\ \bibinfo {pages} {230402} (\bibinfo {year}
  {2011})}\BibitemShut {NoStop}%
\bibitem [{\citenamefont {Vidanovi{\'c}}\ \emph {et~al.}(2013)\citenamefont
  {Vidanovi{\'c}}, \citenamefont {van Druten},\ and\ \citenamefont
  {Haque}}]{vidanovic2013spin}%
  \BibitemOpen
  \bibfield  {author} {\bibinfo {author} {\bibfnamefont {I.}~\bibnamefont
  {Vidanovi{\'c}}}, \bibinfo {author} {\bibfnamefont {N.~J.}\ \bibnamefont {van
  Druten}}, \ and\ \bibinfo {author} {\bibfnamefont {M.}~\bibnamefont
  {Haque}},\ }\href {http://stacks.iop.org/1367-2630/15/i=3/a=035008}
  {\bibfield  {journal} {\bibinfo  {journal} {New Journal of Physics}\ }\textbf
  {\bibinfo {volume} {15}},\ \bibinfo {pages} {035008} (\bibinfo {year}
  {2013})}\BibitemShut {NoStop}%
\bibitem [{\citenamefont {Gautam}\ and\ \citenamefont
  {Angom}(2011)}]{gautam2011phase}%
  \BibitemOpen
  \bibfield  {author} {\bibinfo {author} {\bibfnamefont {S.}~\bibnamefont
  {Gautam}}\ and\ \bibinfo {author} {\bibfnamefont {D.}~\bibnamefont {Angom}},\
  }\href {http://stacks.iop.org/0953-4075/44/i=2/a=025302} {\bibfield
  {journal} {\bibinfo  {journal} {Journal of Physics B: Atomic, Molecular and
  Optical Physics}\ }\textbf {\bibinfo {volume} {44}},\ \bibinfo {pages}
  {025302} (\bibinfo {year} {2011})}\BibitemShut {NoStop}%
\bibitem [{\citenamefont {Lee}\ \emph {et~al.}(2016)\citenamefont {Lee},
  \citenamefont {J\o{}rgensen}, \citenamefont {Liu}, \citenamefont {Wacker},
  \citenamefont {Arlt},\ and\ \citenamefont {Proukakis}}]{PhysRevA.94.013602}%
  \BibitemOpen
  \bibfield  {author} {\bibinfo {author} {\bibfnamefont {K.~L.}\ \bibnamefont
  {Lee}}, \bibinfo {author} {\bibfnamefont {N.~B.}\ \bibnamefont
  {J\o{}rgensen}}, \bibinfo {author} {\bibfnamefont {I.-K.}\ \bibnamefont
  {Liu}}, \bibinfo {author} {\bibfnamefont {L.}~\bibnamefont {Wacker}},
  \bibinfo {author} {\bibfnamefont {J.~J.}\ \bibnamefont {Arlt}}, \ and\
  \bibinfo {author} {\bibfnamefont {N.~P.}\ \bibnamefont {Proukakis}},\ }\href
  {\doibase 10.1103/PhysRevA.94.013602} {\bibfield  {journal} {\bibinfo
  {journal} {Phys. Rev. A}\ }\textbf {\bibinfo {volume} {94}},\ \bibinfo
  {pages} {013602} (\bibinfo {year} {2016})}\BibitemShut {NoStop}%
\bibitem [{\citenamefont {Liu}(2009)}]{doi:10.1063/1.3243875}%
  \BibitemOpen
  \bibfield  {author} {\bibinfo {author} {\bibfnamefont {Z.}~\bibnamefont
  {Liu}},\ }\href {\doibase 10.1063/1.3243875} {\bibfield  {journal} {\bibinfo
  {journal} {Journal of Mathematical Physics}\ }\textbf {\bibinfo {volume}
  {50}},\ \bibinfo {pages} {102104} (\bibinfo {year} {2009})},\ \Eprint
  {http://arxiv.org/abs/https://doi.org/10.1063/1.3243875}
  {https://doi.org/10.1063/1.3243875} \BibitemShut {NoStop}%
\bibitem [{\citenamefont {Tojo}\ \emph {et~al.}(2010)\citenamefont {Tojo},
  \citenamefont {Taguchi}, \citenamefont {Masuyama}, \citenamefont {Hayashi},
  \citenamefont {Saito},\ and\ \citenamefont {Hirano}}]{PhysRevA.82.033609}%
  \BibitemOpen
  \bibfield  {author} {\bibinfo {author} {\bibfnamefont {S.}~\bibnamefont
  {Tojo}}, \bibinfo {author} {\bibfnamefont {Y.}~\bibnamefont {Taguchi}},
  \bibinfo {author} {\bibfnamefont {Y.}~\bibnamefont {Masuyama}}, \bibinfo
  {author} {\bibfnamefont {T.}~\bibnamefont {Hayashi}}, \bibinfo {author}
  {\bibfnamefont {H.}~\bibnamefont {Saito}}, \ and\ \bibinfo {author}
  {\bibfnamefont {T.}~\bibnamefont {Hirano}},\ }\href {\doibase
  10.1103/PhysRevA.82.033609} {\bibfield  {journal} {\bibinfo  {journal} {Phys.
  Rev. A}\ }\textbf {\bibinfo {volume} {82}},\ \bibinfo {pages} {033609}
  (\bibinfo {year} {2010})}\BibitemShut {NoStop}%
\bibitem [{\citenamefont {Zhu}\ and\ \citenamefont
  {Li}(2017)}]{doi:10.1142/S0217984917502153}%
  \BibitemOpen
  \bibfield  {author} {\bibinfo {author} {\bibfnamefont {L.}~\bibnamefont
  {Zhu}}\ and\ \bibinfo {author} {\bibfnamefont {J.}~\bibnamefont {Li}},\
  }\href {\doibase 10.1142/S0217984917502153} {\bibfield  {journal} {\bibinfo
  {journal} {Modern Physics Letters B}\ }\textbf {\bibinfo {volume} {31}},\
  \bibinfo {pages} {1750215} (\bibinfo {year} {2017})},\ \Eprint
  {http://arxiv.org/abs/https://doi.org/10.1142/S0217984917502153}
  {https://doi.org/10.1142/S0217984917502153} \BibitemShut {NoStop}%
\bibitem [{\citenamefont {Wen}\ \emph {et~al.}(2012)\citenamefont {Wen},
  \citenamefont {Liu}, \citenamefont {Cai}, \citenamefont {Zhang},\ and\
  \citenamefont {Hu}}]{PhysRevA.85.043602}%
  \BibitemOpen
  \bibfield  {author} {\bibinfo {author} {\bibfnamefont {L.}~\bibnamefont
  {Wen}}, \bibinfo {author} {\bibfnamefont {W.~M.}\ \bibnamefont {Liu}},
  \bibinfo {author} {\bibfnamefont {Y.}~\bibnamefont {Cai}}, \bibinfo {author}
  {\bibfnamefont {J.~M.}\ \bibnamefont {Zhang}}, \ and\ \bibinfo {author}
  {\bibfnamefont {J.}~\bibnamefont {Hu}},\ }\href {\doibase
  10.1103/PhysRevA.85.043602} {\bibfield  {journal} {\bibinfo  {journal} {Phys.
  Rev. A}\ }\textbf {\bibinfo {volume} {85}},\ \bibinfo {pages} {043602}
  (\bibinfo {year} {2012})}\BibitemShut {NoStop}%
\bibitem [{\citenamefont {Xi}\ \emph {et~al.}(2011)\citenamefont {Xi},
  \citenamefont {Li},\ and\ \citenamefont {Shi}}]{PhysRevA.84.013619}%
  \BibitemOpen
  \bibfield  {author} {\bibinfo {author} {\bibfnamefont {K.-T.}\ \bibnamefont
  {Xi}}, \bibinfo {author} {\bibfnamefont {J.}~\bibnamefont {Li}}, \ and\
  \bibinfo {author} {\bibfnamefont {D.-N.}\ \bibnamefont {Shi}},\ }\href
  {\doibase 10.1103/PhysRevA.84.013619} {\bibfield  {journal} {\bibinfo
  {journal} {Phys. Rev. A}\ }\textbf {\bibinfo {volume} {84}},\ \bibinfo
  {pages} {013619} (\bibinfo {year} {2011})}\BibitemShut {NoStop}%
\bibitem [{\citenamefont {Bandyopadhyay}\ \emph {et~al.}(2017)\citenamefont
  {Bandyopadhyay}, \citenamefont {Roy},\ and\ \citenamefont
  {Angom}}]{PhysRevA.96.043603}%
  \BibitemOpen
  \bibfield  {author} {\bibinfo {author} {\bibfnamefont {S.}~\bibnamefont
  {Bandyopadhyay}}, \bibinfo {author} {\bibfnamefont {A.}~\bibnamefont {Roy}},
  \ and\ \bibinfo {author} {\bibfnamefont {D.}~\bibnamefont {Angom}},\ }\href
  {\doibase 10.1103/PhysRevA.96.043603} {\bibfield  {journal} {\bibinfo
  {journal} {Phys. Rev. A}\ }\textbf {\bibinfo {volume} {96}},\ \bibinfo
  {pages} {043603} (\bibinfo {year} {2017})}\BibitemShut {NoStop}%
\bibitem [{\citenamefont {Wang}\ \emph {et~al.}(2010)\citenamefont {Wang},
  \citenamefont {Gao}, \citenamefont {Jian},\ and\ \citenamefont
  {Zhai}}]{PhysRevLett.105.160403}%
  \BibitemOpen
  \bibfield  {author} {\bibinfo {author} {\bibfnamefont {C.}~\bibnamefont
  {Wang}}, \bibinfo {author} {\bibfnamefont {C.}~\bibnamefont {Gao}}, \bibinfo
  {author} {\bibfnamefont {C.-M.}\ \bibnamefont {Jian}}, \ and\ \bibinfo
  {author} {\bibfnamefont {H.}~\bibnamefont {Zhai}},\ }\href {\doibase
  10.1103/PhysRevLett.105.160403} {\bibfield  {journal} {\bibinfo  {journal}
  {Phys. Rev. Lett.}\ }\textbf {\bibinfo {volume} {105}},\ \bibinfo {pages}
  {160403} (\bibinfo {year} {2010})}\BibitemShut {NoStop}%
\bibitem [{\citenamefont {Gautam}\ and\ \citenamefont
  {Adhikari}(2014)}]{gautam2014phase}%
  \BibitemOpen
  \bibfield  {author} {\bibinfo {author} {\bibfnamefont {S.}~\bibnamefont
  {Gautam}}\ and\ \bibinfo {author} {\bibfnamefont {S.~K.}\ \bibnamefont
  {Adhikari}},\ }\href {\doibase 10.1103/PhysRevA.90.043619} {\bibfield
  {journal} {\bibinfo  {journal} {Phys. Rev. A}\ }\textbf {\bibinfo {volume}
  {90}},\ \bibinfo {pages} {043619} (\bibinfo {year} {2014})}\BibitemShut
  {NoStop}%
\bibitem [{\citenamefont {Li}\ \emph {et~al.}(2018)\citenamefont {Li},
  \citenamefont {Yu}, \citenamefont {Jiang},\ and\ \citenamefont
  {Liu}}]{li2018phase}%
  \BibitemOpen
  \bibfield  {author} {\bibinfo {author} {\bibfnamefont {J.}~\bibnamefont
  {Li}}, \bibinfo {author} {\bibfnamefont {Y.-M.}\ \bibnamefont {Yu}}, \bibinfo
  {author} {\bibfnamefont {K.-J.}\ \bibnamefont {Jiang}}, \ and\ \bibinfo
  {author} {\bibfnamefont {W.-M.}\ \bibnamefont {Liu}},\ }\href@noop {}
  {\bibfield  {journal} {\bibinfo  {journal} {arXiv preprint arXiv:1802.00138}\
  } (\bibinfo {year} {2018})}\BibitemShut {NoStop}%
\bibitem [{\citenamefont {Bhuvaneswari}\ \emph {et~al.}(2018)\citenamefont
  {Bhuvaneswari}, \citenamefont {Nithyanandan},\ and\ \citenamefont
  {Muruganandam}}]{2399-6528-2-2-025008}%
  \BibitemOpen
  \bibfield  {author} {\bibinfo {author} {\bibfnamefont {S.}~\bibnamefont
  {Bhuvaneswari}}, \bibinfo {author} {\bibfnamefont {K.}~\bibnamefont
  {Nithyanandan}}, \ and\ \bibinfo {author} {\bibfnamefont {P.}~\bibnamefont
  {Muruganandam}},\ }\href {http://stacks.iop.org/2399-6528/2/i=2/a=025008}
  {\bibfield  {journal} {\bibinfo  {journal} {Journal of Physics
  Communications}\ }\textbf {\bibinfo {volume} {2}},\ \bibinfo {pages} {025008}
  (\bibinfo {year} {2018})}\BibitemShut {NoStop}%
\bibitem [{\citenamefont {Stamper-Kurn}\ \emph {et~al.}(1998)\citenamefont
  {Stamper-Kurn}, \citenamefont {Andrews}, \citenamefont {Chikkatur},
  \citenamefont {Inouye}, \citenamefont {Miesner}, \citenamefont {Stenger},\
  and\ \citenamefont {Ketterle}}]{PhysRevLett.80.2027}%
  \BibitemOpen
  \bibfield  {author} {\bibinfo {author} {\bibfnamefont {D.~M.}\ \bibnamefont
  {Stamper-Kurn}}, \bibinfo {author} {\bibfnamefont {M.~R.}\ \bibnamefont
  {Andrews}}, \bibinfo {author} {\bibfnamefont {A.~P.}\ \bibnamefont
  {Chikkatur}}, \bibinfo {author} {\bibfnamefont {S.}~\bibnamefont {Inouye}},
  \bibinfo {author} {\bibfnamefont {H.-J.}\ \bibnamefont {Miesner}}, \bibinfo
  {author} {\bibfnamefont {J.}~\bibnamefont {Stenger}}, \ and\ \bibinfo
  {author} {\bibfnamefont {W.}~\bibnamefont {Ketterle}},\ }\href {\doibase
  10.1103/PhysRevLett.80.2027} {\bibfield  {journal} {\bibinfo  {journal}
  {Phys. Rev. Lett.}\ }\textbf {\bibinfo {volume} {80}},\ \bibinfo {pages}
  {2027} (\bibinfo {year} {1998})}\BibitemShut {NoStop}%
\bibitem [{\citenamefont {Ho}(1998)}]{PhysRevLett.81.742}%
  \BibitemOpen
  \bibfield  {author} {\bibinfo {author} {\bibfnamefont {T.-L.}\ \bibnamefont
  {Ho}},\ }\href {\doibase 10.1103/PhysRevLett.81.742} {\bibfield  {journal}
  {\bibinfo  {journal} {Phys. Rev. Lett.}\ }\textbf {\bibinfo {volume} {81}},\
  \bibinfo {pages} {742} (\bibinfo {year} {1998})}\BibitemShut {NoStop}%
\bibitem [{\citenamefont {Barrett}\ \emph {et~al.}(2001)\citenamefont
  {Barrett}, \citenamefont {Sauer},\ and\ \citenamefont
  {Chapman}}]{PhysRevLett.87.010404}%
  \BibitemOpen
  \bibfield  {author} {\bibinfo {author} {\bibfnamefont {M.~D.}\ \bibnamefont
  {Barrett}}, \bibinfo {author} {\bibfnamefont {J.~A.}\ \bibnamefont {Sauer}},
  \ and\ \bibinfo {author} {\bibfnamefont {M.~S.}\ \bibnamefont {Chapman}},\
  }\href {\doibase 10.1103/PhysRevLett.87.010404} {\bibfield  {journal}
  {\bibinfo  {journal} {Phys. Rev. Lett.}\ }\textbf {\bibinfo {volume} {87}},\
  \bibinfo {pages} {010404} (\bibinfo {year} {2001})}\BibitemShut {NoStop}%
\bibitem [{\citenamefont {Kawaguchi}\ and\ \citenamefont
  {Ueda}(2012)}]{KAWAGUCHI2012253}%
  \BibitemOpen
  \bibfield  {author} {\bibinfo {author} {\bibfnamefont {Y.}~\bibnamefont
  {Kawaguchi}}\ and\ \bibinfo {author} {\bibfnamefont {M.}~\bibnamefont
  {Ueda}},\ }\href {\doibase https://doi.org/10.1016/j.physrep.2012.07.005}
  {\bibfield  {journal} {\bibinfo  {journal} {Physics Reports}\ }\textbf
  {\bibinfo {volume} {520}},\ \bibinfo {pages} {253 } (\bibinfo {year}
  {2012})},\ \bibinfo {note} {spinor Bose--Einstein condensates}\BibitemShut
  {NoStop}%
\bibitem [{\citenamefont {Proukakis}(2013)}]{proukakis2013quantum}%
  \BibitemOpen
  \bibfield  {author} {\bibinfo {author} {\bibfnamefont {N.}~\bibnamefont
  {Proukakis}},\ }\href@noop {} {\emph {\bibinfo {title} {Quantum gases: finite
  temperature and non-equilibrium dynamics}}},\ Vol.~\bibinfo {volume} {1}\
  (\bibinfo  {publisher} {World Scientific},\ \bibinfo {year}
  {2013})\BibitemShut {NoStop}%
\bibitem [{\citenamefont {Moerdijk}\ \emph {et~al.}(1995)\citenamefont
  {Moerdijk}, \citenamefont {Verhaar},\ and\ \citenamefont
  {Axelsson}}]{PhysRevA.51.4852}%
  \BibitemOpen
  \bibfield  {author} {\bibinfo {author} {\bibfnamefont {A.~J.}\ \bibnamefont
  {Moerdijk}}, \bibinfo {author} {\bibfnamefont {B.~J.}\ \bibnamefont
  {Verhaar}}, \ and\ \bibinfo {author} {\bibfnamefont {A.}~\bibnamefont
  {Axelsson}},\ }\href {\doibase 10.1103/PhysRevA.51.4852} {\bibfield
  {journal} {\bibinfo  {journal} {Phys. Rev. A}\ }\textbf {\bibinfo {volume}
  {51}},\ \bibinfo {pages} {4852} (\bibinfo {year} {1995})}\BibitemShut
  {NoStop}%
\bibitem [{\citenamefont {Inouye}\ \emph {et~al.}(1998)\citenamefont {Inouye},
  \citenamefont {Andrews}, \citenamefont {Stenger}, \citenamefont {Miesner},
  \citenamefont {Stamper-Kurn},\ and\ \citenamefont
  {Ketterle}}]{inouye1998observation}%
  \BibitemOpen
  \bibfield  {author} {\bibinfo {author} {\bibfnamefont {S.}~\bibnamefont
  {Inouye}}, \bibinfo {author} {\bibfnamefont {M.}~\bibnamefont {Andrews}},
  \bibinfo {author} {\bibfnamefont {J.}~\bibnamefont {Stenger}}, \bibinfo
  {author} {\bibfnamefont {H.-J.}\ \bibnamefont {Miesner}}, \bibinfo {author}
  {\bibfnamefont {D.}~\bibnamefont {Stamper-Kurn}}, \ and\ \bibinfo {author}
  {\bibfnamefont {W.}~\bibnamefont {Ketterle}},\ }\href@noop {} {\bibfield
  {journal} {\bibinfo  {journal} {Nature}\ }\textbf {\bibinfo {volume} {392}},\
  \bibinfo {pages} {151} (\bibinfo {year} {1998})}\BibitemShut {NoStop}%
\bibitem [{\citenamefont {Chin}\ \emph {et~al.}(2010)\citenamefont {Chin},
  \citenamefont {Grimm}, \citenamefont {Julienne},\ and\ \citenamefont
  {Tiesinga}}]{RevModPhys.82.1225}%
  \BibitemOpen
  \bibfield  {author} {\bibinfo {author} {\bibfnamefont {C.}~\bibnamefont
  {Chin}}, \bibinfo {author} {\bibfnamefont {R.}~\bibnamefont {Grimm}},
  \bibinfo {author} {\bibfnamefont {P.}~\bibnamefont {Julienne}}, \ and\
  \bibinfo {author} {\bibfnamefont {E.}~\bibnamefont {Tiesinga}},\ }\href
  {\doibase 10.1103/RevModPhys.82.1225} {\bibfield  {journal} {\bibinfo
  {journal} {Rev. Mod. Phys.}\ }\textbf {\bibinfo {volume} {82}},\ \bibinfo
  {pages} {1225} (\bibinfo {year} {2010})}\BibitemShut {NoStop}%
\bibitem [{\citenamefont {Frapolli}\ \emph {et~al.}(2017)\citenamefont
  {Frapolli}, \citenamefont {Zibold}, \citenamefont {Invernizzi}, \citenamefont
  {Jim\'enez-Garc\'{\i}a}, \citenamefont {Dalibard},\ and\ \citenamefont
  {Gerbier}}]{PhysRevLett.119.050404}%
  \BibitemOpen
  \bibfield  {author} {\bibinfo {author} {\bibfnamefont {C.}~\bibnamefont
  {Frapolli}}, \bibinfo {author} {\bibfnamefont {T.}~\bibnamefont {Zibold}},
  \bibinfo {author} {\bibfnamefont {A.}~\bibnamefont {Invernizzi}}, \bibinfo
  {author} {\bibfnamefont {K.}~\bibnamefont {Jim\'enez-Garc\'{\i}a}}, \bibinfo
  {author} {\bibfnamefont {J.}~\bibnamefont {Dalibard}}, \ and\ \bibinfo
  {author} {\bibfnamefont {F.}~\bibnamefont {Gerbier}},\ }\href {\doibase
  10.1103/PhysRevLett.119.050404} {\bibfield  {journal} {\bibinfo  {journal}
  {Phys. Rev. Lett.}\ }\textbf {\bibinfo {volume} {119}},\ \bibinfo {pages}
  {050404} (\bibinfo {year} {2017})}\BibitemShut {NoStop}%
\end{thebibliography}%




\end{document}